\documentclass[journal,draftcls,onecolumn,12pt,twoside]{IEEEtran}
\normalsize
\usepackage{subcaption}
 \usepackage{setspace}
 \usepackage[font=small,skip=0pt]{caption}

\ifCLASSINFOpdf
\else
\fi

\ifCLASSINFOpdf
\usepackage[pdftex]{graphicx}
\else
\usepackage[dvips]{graphicx}
\fi
\usepackage{url}

\usepackage[cmex10]{amsmath}
\makeatletter
\g@addto@macro\normalsize{%
	\setlength\abovedisplayskip{2pt}
	\setlength\belowdisplayskip{2pt}
	\setlength\abovedisplayshortskip{2pt}
	\setlength\belowdisplayshortskip{2pt}
}
\makeatother
\usepackage{amssymb}
\usepackage{color}

\allowdisplaybreaks

\usepackage{array}

\usepackage{cite}
\usepackage{bbm}

\usepackage{amsmath}
\usepackage{accents}
\newlength{\dhatheight}
\newcommand{\doublehat}[1]{%
    \settoheight{\dhatheight}{\ensuremath{\hat{#1}}}%
    \addtolength{\dhatheight}{-0.35ex}%
    \hat{\vphantom{\rule{1pt}{\dhatheight}}%
    \smash{\hat{#1}}}}

\newtheorem{theorem}{Theorem}
\newtheorem{definition}{Definition}
\newtheorem{lemma}{Lemma}

\newtheorem{corollary}{Corollary}
\newtheorem{remark}{Remark}
\usepackage[text={6.5in,9.5in},centering]{geometry}
\usepackage{graphicx}

\begin{document}

%
\title{On Distributed Lossy Coding of Symmetrically Correlated Gaussian Sources}

\author{Siyao~Zhou, Sadaf~Salehkalaibar, Jingjing~Qian, Jun~Chen, Wuxian~Shi, Yiqun~Ge, and Wen~Tong
	\thanks{Siyao Zhou, Sadaf Salehkalaibar, Jingjing Qian, and Jun Chen are with the Department of Electrical and Computer Engineering at McMaster University, Hamilton, ON L8S 4K1, Canada (email: \{zhous58, qianj40, salehkas, chenjun\}@mcmaster.ca).}
	\thanks{Wuxian Shi, Yiqun Ge, and Wen Tong are with   the Ottawa Research Center, Huawei Technologies, Ottawa, ON K2K 3J1, Canada (email: \{wuxian.shi, yiqun.ge, tongwen\}@huawei.com)}
}

\maketitle

\begin{abstract}
A distributed lossy compression network with $L$ encoders and a decoder is considered.
Each encoder observes a source and sends a compressed version to the decoder. The decoder produces a joint reconstruction of target signals with the mean squared error distortion below a given threshold. It is assumed that the observed sources can be expressed as the sum of target signals and corruptive noises which are independently generated from two symmetric multivariate Gaussian distributions. The minimum compression rate of this network versus the distortion threshold is referred to as the  rate-distortion function, for which an explicit lower bound is established by solving a minimization problem. Our lower bound matches the well-known Berger-Tung upper bound for some values of the distortion threshold. The asymptotic gap between the upper and lower bounds is characterized in the large $L$ limit.


\end{abstract}

\IEEEpeerreviewmaketitle

\begin{IEEEkeywords}
Asymptotic analysis, distributed source coding, rate-distortion function, vector Gaussian source.
\end{IEEEkeywords}

\section{Introduction}

Recently, there has been an increase in the deployment of sensor applications in wireless networks as parts of the future Internet of Things (IoT), thanks to the decreasing cost of sensors. One of the theoretical challenges that arises in these systems is to reduce the amount of data that is transmitted in the network by processing it locally at each sensor. A possible solution to this problem is to exploit the statistical dependency among the data at different sensors to get an improved compression efficiency. The multi-terminal source coding theory aims to develop suitable schemes for that purpose and characterize the corresponding performance limits. There have been significant amount of works over the past few decades in this area, e.g., Slepian-Wolf source coding \cite{SW} for lossless compression, more recent works on Gaussian multi-terminal source coding and its variants \cite{Oohama1, Oohama2, Prabhak, Jun, Jun2, Wagner, wang2010sum, wang2013vector,wang2014vector, Oohama4, Jun4, Oohama3}. An interesting regime that has received particular attention (see, e.g., \cite{Oohama2}) is when the number of encoders in the network approaches infinity. This asymptotic regime reflects the typical scenarios in sensor fusion and is also relevant to some emerging machine learning applications (esp., federated learning) that leverage distributed compression to reduce the communication cost between the central server and a massive number of edge devices for training a global model.

In the present paper, we study a  compression system with $L$ distributed encoders and a central decoder. Each encoder compresses its observed source sequence and forwards the compressed version to the decoder. The decoder is required to reconstruct the target signals with the mean squared error distortion below a given threshold. It is assumed that the observed sources can be expressed as the sum of target signals and corruptive noises which are generated independently according to two symmetric multivariate Gaussian distributions.  
 We are interested in characterizing the minimum required compression rate as a function of the distortion threshold, which is known as the rate-distortion function. Our setup is different from the  Gaussian CEO problem \cite{CEO} in two aspects. Firstly,  the target signals are assumed to form a vector process. Secondly, the noises across different encoders are allowed to be correlated with each other. Notice that these two relaxations do not exist in the original Gaussian CEO problem where the target signal is a scalar process and the noises across different encoders are independent. A generalized version of the Gaussian CEO problem that allows the noises to be symmetrically correlated across different encoders is considered in \cite{Jun}, which establishes, among others, a lower bound on the rate-distortion function. Unfortunately, this lower bound is given in the form of a non-trivial minimization  program and consequently is not amenable to direct  analytical/numerical evaluation. 

As a main contribution of this work, we derive a closed form expression of this lower bound by solving the minimization program explicitly and make a systematic comparison with  the well-known Berger-Tung upper bound \cite[Thm 12.1]{ElGamal}.  It should be mentioned that the symmetry assumption adopted in our setup is not critical for our analysis. It only helps us to present the rate-distortion expressions in explicit forms. We also provide an asymptotic analysis of the upper and lower bounds in the large $L$ limit, extending Oohama's celebrated result \cite{Oohama2} for the Gaussian CEO problem.


The rest of this paper is organized as follows. The system model and some preliminaries are presented in Section~\ref{sec:sys-model}. The main results are stated in Section~\ref{sec:main-result} while their proofs are given in Sections~\ref{conv-new-proof},~\ref{upper-asym-proof} and~\ref{lower-asym-thm-proof}. The paper is concluded in Section~\ref{sec:conclusion}.

\subsection{Notation}\label{sec2}
We basically follow the notation in \cite{Jun}. Specifically, $\mathbb{E}[.]$, $(.)^T$, $\text{tr}(.)$ and $\text{det}(.)$ represent the expectation, transpose, trace and determinant operators, respectively. An $L\times L$ diagonal matrix with diagonal entries $\kappa_1,\ldots, \kappa_L$ is denoted $\text{diag}^{(L)}(\kappa_1,\ldots,\kappa_L)$. An $L$-dimensional all-one row vector is written as $\textbf{1}_L$. We use
$X^n$ as an abbreviation of  $(X_1,\ldots,X_n)$. For a set $\mathcal{A}$ with elements $a_1<\ldots<a_L$,  $(\omega_{a_\ell})_{\ell\in\mathcal{A}}$ means  $(\omega_{a_1},\ldots,\omega_{a_L})$. The cardinality of a set $\mathcal{X}$ is denoted $|\mathcal{X}|$. In this paper, the base of the logarithm function is $e$.

\section{System Model}\label{sec:sys-model}
  
Consider a multi-terminal source coding problem with $L$ distributed encoders and a centralized decoder. There are $L$ sources $(X_1,\ldots,X_{L})\in \mathbb{R}^L$, which form a zero-mean Gaussian vector. The encoders observe the noisy versions of these sources, denoted by $(Y_1,\ldots,Y_L)\in \mathbb{R}^L$, which can be expressed as
\begin{IEEEeqnarray}{rCl}
Y_{\ell} = X_{\ell} + Z_{\ell},\qquad \ell\in \{1,\ldots,L\},
\end{IEEEeqnarray}
where $(Z_1,\ldots, Z_{L})$ is a zero-mean Gaussian random vector independent of $(X_1,\ldots, X_{L})$. We define  $\textbf{X}:= (X_1,\ldots,$ $X_L)^T$, $\textbf{Y}:= (Y_1,\ldots, Y_L)^T$, and $\textbf{Z}:= (Z_1,\ldots,Z_L)^T$. The distributions of $\textbf{X}$, $\textbf{Y}$ and $\textbf{Z}$ are  determined by their covariance matrices $\Sigma_X$, $\Sigma_Y$ and $\Sigma_Z$, respectively.

The source vector $\textbf{X}$ together with the noise vector $\textbf{Z}$ and the corrupted version $\textbf{Y}$ generates an i.i.d. process $\{(\textbf{X}_i,\textbf{Y}_i,\textbf{Z}_i)\}$. 
Each encoder $\ell \in \{1,\ldots, L\}$ assigns a message $M_{\ell}\in\mathcal{M}_{\ell}$ to its observed sequence $Y_{\ell}^n$ using an encoding function $\phi_{\ell}^{(n)}\colon \mathbb{R}^n\to \mathcal{M}_{\ell}$ such that $M_{\ell} := \phi_{\ell}^{(n)}(Y_{\ell}^n)$. Given $(M_1,\ldots,M_L)$, the decoder produces a reconstruction 
$(\hat{X}_1^n,\ldots,\hat{X}_{L}^n):=g^{(n)}(M_{1},\ldots,M_{L})$ using a decoding function  $g^{(n)}\colon \mathcal{M}_{1}\times \ldots \mathcal{M}_{L}\to \mathbb{R}^{L\times n}$. 

\begin{definition}\label{model-def} A rate-distortion pair $(R, D)$ is said to be achievable if for any $\epsilon>0$, there exist encoding functions $\phi_{\ell}^{(n)}$, $\ell\in\{1,\cdots,L\}$, and a decoding function $g^{(n)}$ such that
\begin{IEEEeqnarray}{rCl}
\frac{1}{n}\sum_{\ell=1}^L\log |\mathcal{M}_{\ell}|\leq R+\epsilon,\label{rate-cons}
\end{IEEEeqnarray}
and 
\begin{IEEEeqnarray}{rCl}
\frac{1}{nL}\sum_{\ell=1}^L\sum_{i=1}^n \mathbb{E}[(X_{\ell,i}-\hat{X}_{\ell,i})^2]\leq D+\epsilon.
\end{IEEEeqnarray}
For every $D$, let $\mathcal{R}(D)$ denote the infimum of $R$ such that $(R,D)$ is achievable. We shall refer to 
 $\mathcal{R}(D)$ as the rate-distortion function.
\end{definition}
\subsection{Preliminaries}

For a given $L\times L$ matrix 
\begin{IEEEeqnarray}{rCl}
\Gamma:= \begin{pmatrix}\alpha & \beta & \ldots & \beta\\[-1.5ex] \beta & \alpha & \ldots & \beta\\[-1.5ex]\vdots& \vdots & \ldots & \vdots\\[-1.5ex]\beta & \beta & \ldots & \alpha\end{pmatrix},
\end{IEEEeqnarray}
it follows by the eigenvalue decomposition that we can write
\begin{IEEEeqnarray}{rCl}
\Gamma = \Theta \Lambda \Theta^T,
\end{IEEEeqnarray}
where $\Theta$ is an arbitrary unitary matrix with the first column being $\frac{1}{\sqrt{L}}\textbf{1}_L^T$ and 
\begin{IEEEeqnarray}{rCl}
\Lambda := \text{diag}^{(L)} (\alpha+(L-1)\beta,\alpha-\beta,\ldots,\alpha-\beta).
\end{IEEEeqnarray}
In this work, we assume that the covariance matrix $\Sigma_*$, $*\in \{X,Y,Z\}$, can be written as 
\begin{IEEEeqnarray}{rCl}
\Sigma_*:= \begin{pmatrix}\sigma_*^2 & \rho_*\sigma_*^2 & \ldots & \rho_*\sigma_*^2\\[-1.5ex] \rho_*\sigma_*^2 & \sigma_*^2 & \ldots & \rho_*\sigma_*^2\\[-1.5ex]\vdots& \vdots & \ldots & \vdots\\[-1.5ex]\rho_*\sigma_*^2 & \rho_*\sigma_*^2 & \ldots & \sigma_*^2\end{pmatrix},
\end{IEEEeqnarray}
for some  $\sigma_*$ and $\rho_*$. Therefore, we can write
\begin{IEEEeqnarray}{rCl}
\Sigma_* &= & \Theta\Lambda_*\Theta^T, \label{Sig-decomp}
\end{IEEEeqnarray}
where
\begin{IEEEeqnarray}{rCl}
\Lambda_*&:=&\text{diag}^{(L)}(\lambda_*, \gamma_*,\ldots, \gamma_*)\label{Lam-decomp}
\end{IEEEeqnarray}
with
\begin{subequations}\label{la-ga}
\begin{IEEEeqnarray}{rCl}
\lambda_*& :=& (1+(L-1)\rho_*)\sigma_*^2, \label{lamy-def}\\
\gamma_*& :=& (1-\rho_*)\sigma_*^2.
\end{IEEEeqnarray}
\end{subequations}
Note that it suffices to specify $\Sigma_X$ and $\Sigma_Y$ since $\Sigma_Y=\Sigma_X+\Sigma_Z$ (i.e., $\sigma^2_Y=\sigma^2_X+\sigma^2_Z$ and $\rho_Y\sigma^2_Y=\rho_X\sigma^2_X+\rho_Z\sigma^2_Z$). It is also clear that $\lambda_Y=\lambda_X+\lambda_Z$ and $\gamma_Y=\gamma_X+\gamma_Z$.
To ensure that the covariance matrices are positive semi-definite and the source vector $\textbf{X}$ is not deterministic, we assume $\sigma^2_X>0$, $\sigma^2_Z\geq 0$, $\rho_X\in[-\frac{1}{L-1},1]$ and $\rho_Z\in[-\frac{1}{L-1},1]$; we further assume $\Sigma_Y$ is positive definite, i.e., $\min(\lambda_Y,\gamma_Y)>0$.


\section{Main Results}\label{sec:main-result}

First, we review some results of \cite{Jun}. The following theorem gives an upper bound on the rate-distortion function $\mathcal{R}(D)$. Let
\begin{IEEEeqnarray}{rCl}
d_{\min}:= \frac{\lambda_X\lambda_Z}{L\lambda_Y}+\frac{(L-1)\gamma_X \gamma_Z}{L\gamma_Y},
\end{IEEEeqnarray}
and
\begin{IEEEeqnarray}{rCl}
\overline{\mathcal{R}}(D):= \frac{1}{2}\log \left(1+\frac{\lambda_Y}{\lambda_Q}\right)+\frac{L-1}{2}\log \left(1+\frac{\gamma_Y}{\lambda_Q}\right),\label{ach-rate}
\end{IEEEeqnarray}
where $\lambda_Q$ is a positive number satisfying
\begin{IEEEeqnarray}{rCl}
\lambda_X\left(1-\frac{\lambda_X}{\lambda_Y+\lambda_Q}\right)+(L-1)\gamma_X\left(1-\frac{\gamma_X}{\gamma_Y+\lambda_Q}\right)= LD.\label{distortion-ach}
\end{IEEEeqnarray}
\begin{theorem}[Upper bound of Thm 2 in \cite{Jun}]\label{thm-ach} For $D\in (d_{\min},\sigma_X^2)$, we have
\begin{IEEEeqnarray}{rCl}
\mathcal{R}(D)\leq \overline{\mathcal{R}}(D).
\end{IEEEeqnarray}
\end{theorem}

\underline{\textit{Sketch of Proof of Theorem~\ref{thm-ach}:}} See Appendix~\ref{sketch-thm1-proof}.

\begin{remark} It can be observed that $\overline{\mathcal{R}}(D)$, given in~\eqref{ach-rate}, is expressed as the sum of two terms. These two terms correspond to the compression rates required for the larger eigenvalue $\lambda_Y$ and the smaller eigenvalue $\gamma_Y$, respectively. The second term has the coefficient $L-1$, which is consistent with the fact that the eigenvalue $\gamma_Y$ appears $L-1$ times in the diagonal matrix $\Lambda_Y$. A similar observation can be made for the distortion expression as given in~\eqref{distortion-ach}.
\end{remark}

Next, we review a result of \cite{Jun}, which provides a lower bound on the rate-distortion function $\mathcal{R}(D)$ in the form of a minimization program. Define
\begin{IEEEeqnarray}{rCl}
\Omega(\alpha,\beta,\delta) := \frac{1}{2}\log \frac{\lambda_Y^2}{(\lambda_Y-\lambda_W)\alpha+\lambda_Y\lambda_W}+\frac{L-1}{2}\log \frac{\gamma_Y^2}{(\gamma_Y-\lambda_W)\beta+\gamma_Y\lambda_W}+\frac{L}{2}\log \frac{\lambda_W}{\delta},\nonumber\\\label{general-obj-function}
\end{IEEEeqnarray}
where  $\lambda_W=\min(\lambda_Y,\gamma_Y)$. Let $\underline{\mathcal{R}}(D)$ be the solution of the following optimization problem:
\begin{subequations}\label{general-constraints}\begin{IEEEeqnarray}{rCl}
&&\underline{\mathcal{R}}(D):=\min_{\alpha,\beta,\delta}\;\; \Omega(\alpha,\beta,\delta),\\
&& \qquad \qquad
0<\alpha\leq \lambda_Y\label{opt-consa},\\
&& \qquad \qquad 0<\beta\leq \gamma_Y,\\
&& \qquad \qquad 0<\delta,\\
&& \qquad \qquad \delta \leq (\alpha^{-1}+\lambda_W^{-1}-\lambda_Y^{-1})^{-1},\\
&& \qquad \qquad \delta \leq (\beta^{-1}+\lambda_W^{-1}-\gamma_Y^{-1})^{-1},\\
&& \qquad \qquad\lambda_X^2\lambda_Y^{-2}\alpha+\lambda_X-\lambda_X^2\lambda_Y^{-1}+(L-1)(\gamma_X^2\gamma_Y^{-2}\beta+\gamma_X-\gamma_X^2\gamma_Y^{-1})\leq LD.\label{opt-consz}
\end{IEEEeqnarray}
\end{subequations}

\begin{theorem}[Lower bound of Thm 2 in \cite{Jun}]\label{thm-conv} For $D\in (d_{\min},\sigma_X^2)$, we have
\begin{IEEEeqnarray}{rCl}
\mathcal{R}(D)\geq \underline{\mathcal{R}}(D).
\end{IEEEeqnarray}

\end{theorem}

\underline{\textit{Sketch of Proof of Theorem~\ref{thm-conv}:}} See Appendix~\ref{sketch-thm2-proof}.

In the following, we derive the explicit solution of the above program. Define the following rate-distortion expressions:
 
\begin{IEEEeqnarray}{rCl}
\underline{\mathcal{R}}^c(D)=\left\{\begin{array}{ll} \underline{\mathcal{R}}_1^c(D)& \;\text{if}\;\;\; D\leq \mathsf{D}^c_{\text{th}},\\ \underline{\mathcal{R}}_2^c(D)& \;\text{if}\;\;\; D> \mathsf{D}^c_{\text{th}}, \end{array}\right.\label{Rc-def}
\end{IEEEeqnarray}
\begin{IEEEeqnarray}{rCl}
\hat{\underline{\mathcal{R}}}^c(D):= \left\{\begin{array}{ll} \hat{\underline{\mathcal{R}}}_1^c (D)&\;\text{if}\;\;\; D\leq \hat{\mathsf{D}}^c_{\text{th}}, \\\hat{\underline{\mathcal{R}}}_2^c (D)&\;\text{if}\;\;\; D > \hat{\mathsf{D}}^c_{\text{th}},\end{array} \right.\label{Rchat-def}
\end{IEEEeqnarray}
where
\begin{IEEEeqnarray}{rCl}
\underline{\mathcal{R}}_1^c(D)&:=& \frac{L+1}{2}\log \frac{(L+1)\gamma_X^2\gamma_Y^{-1}}{LD-\lambda_X-(L-1)(\gamma_X-\gamma_X^2\gamma_Y^{-1})+\lambda_X^2\lambda_Y^{-2}(\lambda_Y+(\gamma_Y^{-1}-\lambda_Y^{-1})^{-1})}\nonumber\\
&&\hspace{0.5cm}+\frac{1}{2}\log (\lambda_X^2\gamma_X^{-2}(\lambda_Y\gamma_Y^{-1}-1)^{-1})+\frac{L}{2}\log \frac{L-1}{L},\label{R2Dl}\\
\underline{\hat{\mathcal{R}}}_1^c(D)&:=& \nonumber\\&&\hspace{-0.7cm}\frac{2L-1}{2}\log \frac{(2L-1)\lambda_X^2 \lambda_Y^{-1}}{LD-\lambda_X-(L-1)(\gamma_X-\gamma_X^2\gamma_Y^{-1})+\lambda_X^{2}\lambda_Y^{-1}+(L-1)\gamma_X^{2}\gamma_Y^{-2}(\lambda_Y^{-1}-\gamma_Y^{-1})^{-1}}\nonumber\\
&&\hspace{0.5cm}+\frac{L-1}{2}\log (\gamma_X^2\lambda_X^{-2}(\gamma_Y\lambda_Y^{-1}-1)^{-1})+\frac{L}{2}\log \frac{1}{L},\label{R2Dlhat}\\
\underline{\mathcal{R}}^c_2(D)&:=&\frac{L}{2}\log  \frac{(L-1)\gamma_X^2\gamma_Y^{-1}}{LD-\lambda_X-(L-1)(\gamma_X-\gamma_X^2\gamma_Y^{-1})},\label{R1D}\\
\underline{\hat{ \mathcal{R}}}^c_2(D)&:=&\frac{L}{2}\log  \frac{\lambda_X^2\lambda_Y^{-1}}{LD-\lambda_X-(L-1)\gamma_X+\lambda_X^2\lambda_Y^{-1}},\label{R1Dhat}
\end{IEEEeqnarray}
and
\begin{IEEEeqnarray}{rCl}
\mathsf{D}^c_{\text{th}}&:=&\lambda_X^2(\lambda_Y-\gamma_Y)^{-1}+\frac{1}{L}((L-1)\gamma_X^2(\gamma_X^{-1}-\gamma_Y^{-1})+\lambda_X),\\
\hat{\mathsf{D}}^c_{\text{th}}&:=&\gamma_X^2(\gamma_Y-\lambda_Y)^{-1}+\frac{1}{L}((L-1)\gamma_X+\lambda_X^2(\lambda^{-1}_X-\lambda_Y^{-1})).
\end{IEEEeqnarray}
Moreover, define the following parameters:
\begin{IEEEeqnarray}{rCl}
\mu_1&:=&\frac{1}{2}-\frac{1}{2}\sqrt{1-\frac{4L}{L-1}\lambda_X^2\lambda_Y^{-2}\gamma_X^{-2}\gamma_Y^2},\\
\mu_2&:=& \frac{1}{2}+\frac{1}{2}\sqrt{1-\frac{4L}{L-1}\lambda_X^2\lambda_Y^{-2}\gamma_X^{-2}\gamma_Y^2},\\
\nu_1&:=& \frac{1}{2}-\frac{1}{2} \sqrt{1-4L\gamma_X^2\lambda_Y^2\lambda_X^{-2}\gamma_Y^{-2}},\\
\nu_2&:=& \frac{1}{2}+\frac{1}{2} \sqrt{1-4L\gamma_X^2\lambda_Y^2\lambda_X^{-2}\gamma_Y^{-2}},
\end{IEEEeqnarray}
and 
\begin{IEEEeqnarray}{rCl}
\mathsf{D}_{\text{th},1}&:=&\frac{1}{L}\Big( \lambda_X+(L-1)\gamma_X-\lambda_X^2(\lambda_Y-\gamma_Y)^{-1}+(L-1)\gamma_X^2(\lambda_Y-\gamma_Y)^{-1}\nonumber\\&&\hspace{2cm}-\mu_2(L-1)\gamma_X^2\gamma_Y^{-1}(1-\gamma_Y\lambda_Y^{-1})^{-1}+\frac{1}{\mu_2}\lambda_X^2\lambda_Y^{-1}(\lambda_Y\gamma_Y^{-1}-1)^{-1}\Big),\\
\mathsf{D}_{\text{th},2}&:=& \frac{1}{L}\Big(\lambda_X+(L-1)\gamma_X-\lambda_X^2(\lambda_Y-\gamma_Y)^{-1}+(L-1)\gamma_X^2(\lambda_Y-\gamma_Y)^{-1}\nonumber\\&&\hspace{2cm}-\mu_1(L-1)\gamma_X^2\gamma_Y^{-1}(1-\gamma_Y\lambda_Y^{-1})^{-1}+\frac{1}{\mu_1}\lambda_X^2\lambda_Y^{-1}(\lambda_Y\gamma_Y^{-1}-1)^{-1}\Big),\\
\hat{\mathsf{D}}_{\text{th},1}&:=& \frac{1}{L}\Big(\lambda_X+(L-1)\gamma_X-\lambda_X^2(\lambda_Y-\gamma_Y)^{-1}+(L-1)\gamma_X^2(\lambda_Y-\gamma_Y)^{-1}\nonumber\\&&\hspace{2cm}-\frac{1}{\nu_2}(L-1)\gamma_X^2\gamma_Y^{-1}(1-\gamma_Y\lambda_Y^{-1})^{-1}+\nu_2\lambda_X^2\lambda_Y^{-1}(\lambda_Y\gamma_Y^{-1}-1)^{-1}\Big),\\
\hat{\mathsf{D}}_{\text{th},2}&:=& \frac{1}{L}\Big(\lambda_X+(L-1)\gamma_X-\lambda_X^2(\lambda_Y-\gamma_Y)^{-1}+(L-1)\gamma_X^2(\lambda_Y-\gamma_Y)^{-1}\nonumber\\&&\hspace{2cm}-\frac{1}{\nu_1}(L-1)\gamma_X^2\gamma_Y^{-1}(1-\gamma_Y\lambda_Y^{-1})^{-1}+\nu_1\lambda_X^2\lambda_Y^{-1}(\lambda_Y\gamma_Y^{-1}-1)^{-1}\Big).
\end{IEEEeqnarray}

\begin{theorem}[Lower bound]\label{conv-new} The lower bound $\underline{\mathcal{R}}(D)$ is completely characterized as follows. 
\begin{itemize}
    \item  $\lambda_Y\geq \gamma_Y$:
    \begin{enumerate}
        \item If $\lambda_X^2\gamma_Y^2\geq\frac{L-1}{4L}\gamma_X^{2}\lambda_Y^{2}$ or if $\lambda_X^2\gamma_Y^2<\frac{L-1}{4L}\gamma_X^{2}\lambda_Y^{2}$ and $\mu_2\leq \frac{\gamma_Y}{\lambda_Y}$ , then for $D\in (d_{\min},\sigma_X^2)$, we have
        \begin{IEEEeqnarray}{rCl}
        \underline{\mathcal{R}}(D)=\overline{\mathcal{R}}(D).
                \end{IEEEeqnarray}
        \item If $\lambda_X^2\gamma_Y^2<\frac{L-1}{4L}\gamma_X^{2}\lambda_Y^{2}$, $\mu_1\leq \frac{\gamma_Y}{\lambda_Y}$ and $\frac{\gamma_Y}{\lambda_Y}<\mu_2<1$ , then for $D\in (d_{\min},\sigma_X^2)$, we have
            \begin{IEEEeqnarray}{rCl}
            \underline{\mathcal{R}}(D)=\left\{\begin{array}{ll} \overline{\mathcal{R}}(D), &\hspace{0.5cm} D\leq \mathsf{D}_{\text{th},1},\\\underline{\mathcal{R}}^c(D),&\hspace{0.5cm}D>\mathsf{D}_{\text{th},1}.\end{array} \right.
            \end{IEEEeqnarray}
            \item If $\lambda_X^2\gamma_Y^2<\frac{L-1}{4L}\gamma_X^{2}\lambda_Y^{2}$, $\mu_1>\frac{\gamma_Y}{\lambda_Y}$ and $\mu_2<1$, then for $D\in (d_{\min},\sigma_X^2)$, we have 
            \begin{IEEEeqnarray}{rCl}
             \underline{\mathcal{R}}(D)=\left\{\begin{array}{ll} \overline{\mathcal{R}}(D),&\hspace{0.5cm}  D\leq \mathsf{D}_{\text{th},1},\\\underline{\mathcal{R}}^c(D),&\hspace{0.5cm}\mathsf{D}_{\text{th},1}<D<\mathsf{D}_{\text{th},2},\\\overline{\mathcal{R}}(D),&\hspace{0.5cm}D\geq\mathsf{D}_{\text{th},2}.\end{array}\right.
            \end{IEEEeqnarray}
            \item If $\lambda_X^2\gamma_Y^2<\frac{L-1}{4L}\gamma_X^{2}\lambda_Y^{2}$, $\mu_1=0$ and $\mu_2=1$ (or equivalently, if $\lambda_X=0$), then for $D\in (d_{\min},\sigma_X^2)$, we have
            \begin{IEEEeqnarray}{rCl}
            \underline{\mathcal{R}}(D)=\underline{\mathcal{R}}^c(D).
            \end{IEEEeqnarray}
    \end{enumerate}

\item  $\gamma_Y\geq \lambda_Y$:
\begin{enumerate}
        \item If  $\gamma_X^2\lambda_Y^2\geq\frac{1}{4L}\lambda_X^{2}\gamma_Y^{2}$ or if $\gamma_X^2\lambda_Y^2<\frac{1}{4L}\lambda_X^{2}\gamma_Y^{2}$ and $\nu_2\leq \frac{\lambda_Y}{\gamma_Y}$, then for $D\in (d_{\min},\sigma_X^2)$, we have
        \begin{IEEEeqnarray}{rCl}
        \underline{\mathcal{R}}(D)=\overline{\mathcal{R}}(D).
                \end{IEEEeqnarray}
        \item If $\gamma_X^2\lambda_Y^2<\frac{1}{4L}\lambda_X^{2}\gamma_Y^{2}$, $\nu_1\leq \frac{\lambda_Y}{\gamma_Y}$ and $\frac{\lambda_Y}{\gamma_Y}<\nu_2<1$, then for $D\in (d_{\min},\sigma_X^2)$,  we have
            \begin{IEEEeqnarray}{rCl}
            \underline{\mathcal{R}}(D)=\left\{\begin{array}{ll} \overline{\mathcal{R}}(D), &\hspace{0.5cm} D\leq \hat{\mathsf{D}}_{\text{th},1},\\\hat{\underline{\mathcal{R}}}^c(D),&\hspace{0.5cm}D>\hat{\mathsf{D}}_{\text{th},1}.\end{array} \right.
            \end{IEEEeqnarray}
            \item If $\gamma_X^2\lambda_Y^2<\frac{1}{4L}\lambda_X^{2}\gamma_Y^{2}$, $\nu_1>\frac{\lambda_Y}{\gamma_Y}$ and $\nu_2<1$, then for $D\in (d_{\min},\sigma_X^2)$, we have 
            \begin{IEEEeqnarray}{rCl}
             \underline{\mathcal{R}}(D)=\left\{\begin{array}{ll} \overline{\mathcal{R}}(D),&\hspace{0.5cm}  D\leq \hat{\mathsf{D}}_{\text{th},1},\\\hat{\underline{\mathcal{R}}}^c(D),&\hspace{0.5cm}\hat{\mathsf{D}}_{\text{th},1}<D<\hat{\mathsf{D}}_{\text{th},2},\\\overline{\mathcal{R}}(D),&\hspace{0.5cm}D\geq\hat{\mathsf{D}}_{\text{th},2}.\end{array}\right. 
            \end{IEEEeqnarray}
            \item If $\gamma_X^2\lambda_Y^2<\frac{1}{4L}\lambda_X^{2}\gamma_Y^{2}$, $\nu_1=0$ and $\nu_2=1 $ (or equivalently, if $\gamma_X=0$), then for $D\in (d_{\min},\sigma_X^2)$, we have 
             \begin{IEEEeqnarray}{rCl}
            \underline{\mathcal{R}}(D)=\underline{\mathcal{R}}^c(D).
            \end{IEEEeqnarray}
    \end{enumerate}
\end{itemize}

\end{theorem}
\begin{IEEEproof} See Section~\ref{conv-new-proof}.
\end{IEEEproof}

According to Theorem~\ref{conv-new}, under some conditions, the lower bound $\underline{\mathcal{R}}(D)$ matches the upper bound $\overline{\mathcal{R}}(D)$.
The gap between the lower and upper bounds will be investigated in the following example for some values of the parameters.

\begin{figure}
\centering
\begin{minipage}{0.5\linewidth}
\centering
\includegraphics[width=0.7\linewidth,height=4cm]{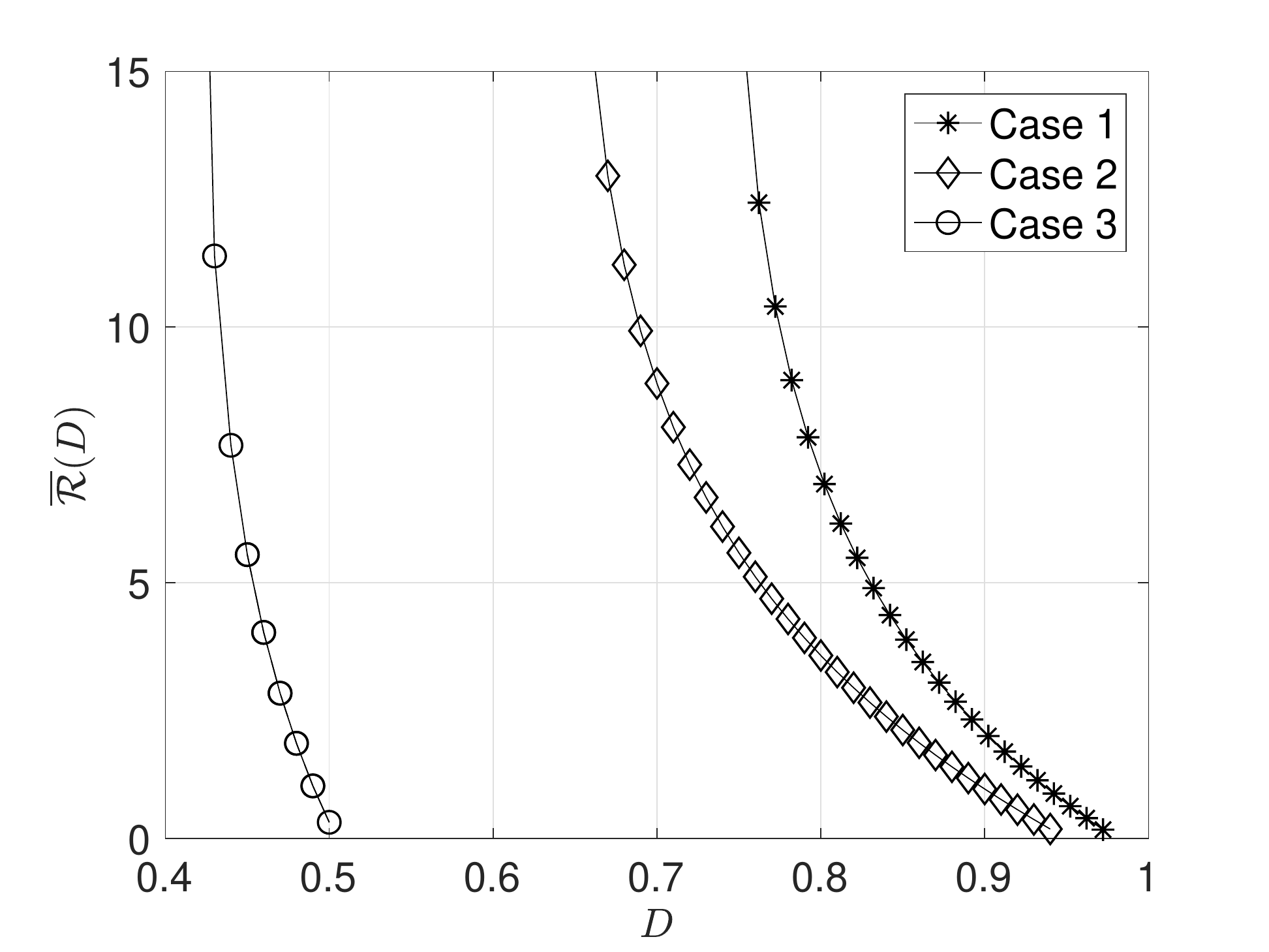}
\subcaption{}
\label{fig:R_UB}
\end{minipage}%
\begin{minipage}{0.5\linewidth}
\centering
\includegraphics[width=0.7\linewidth,height=4cm]{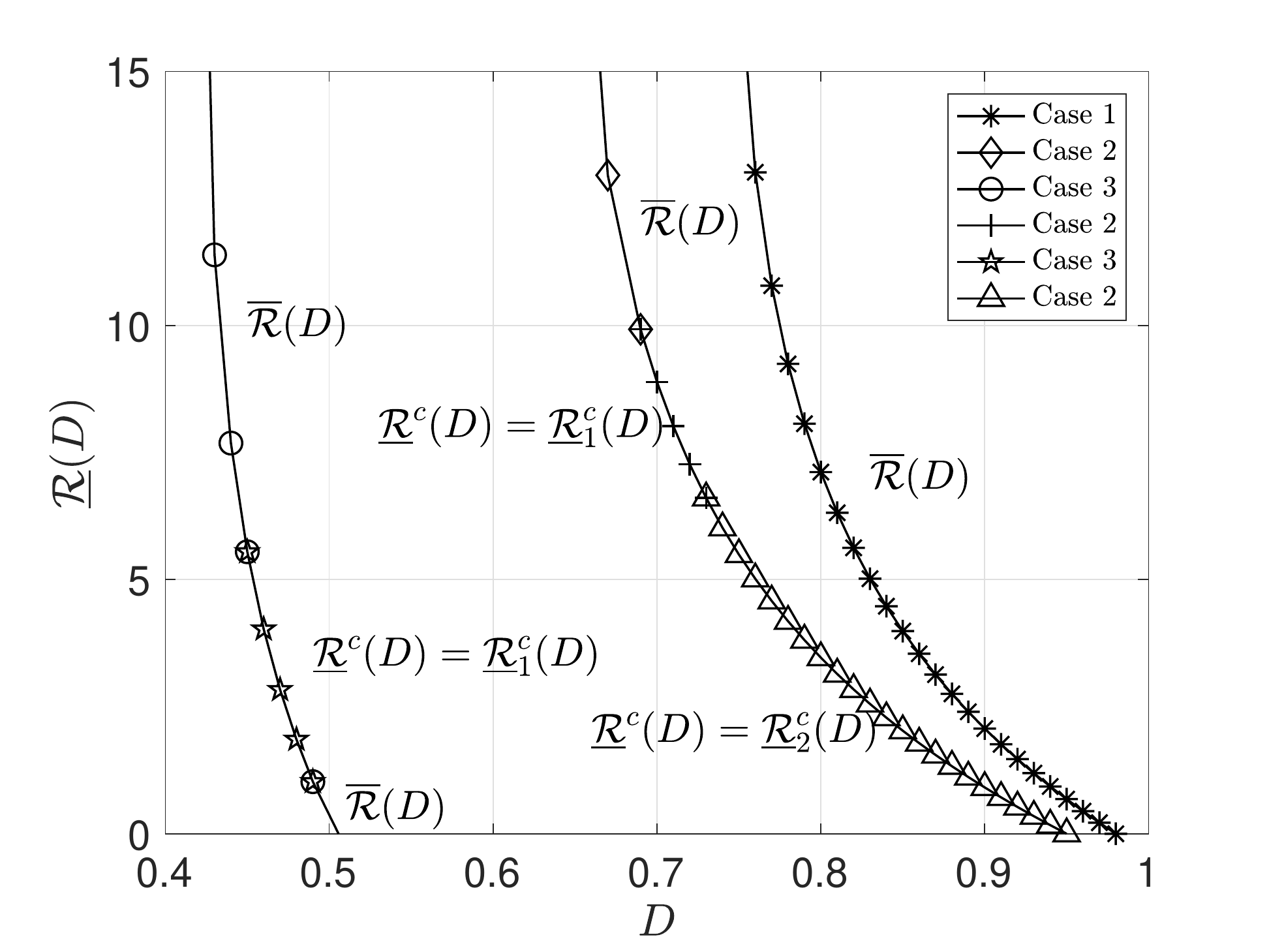}
\subcaption{}
\label{fig:R_LB}
\end{minipage}\\
\centering
\begin{minipage}{0.5\linewidth}
\centering
\includegraphics[width=0.7\linewidth,height=4cm]{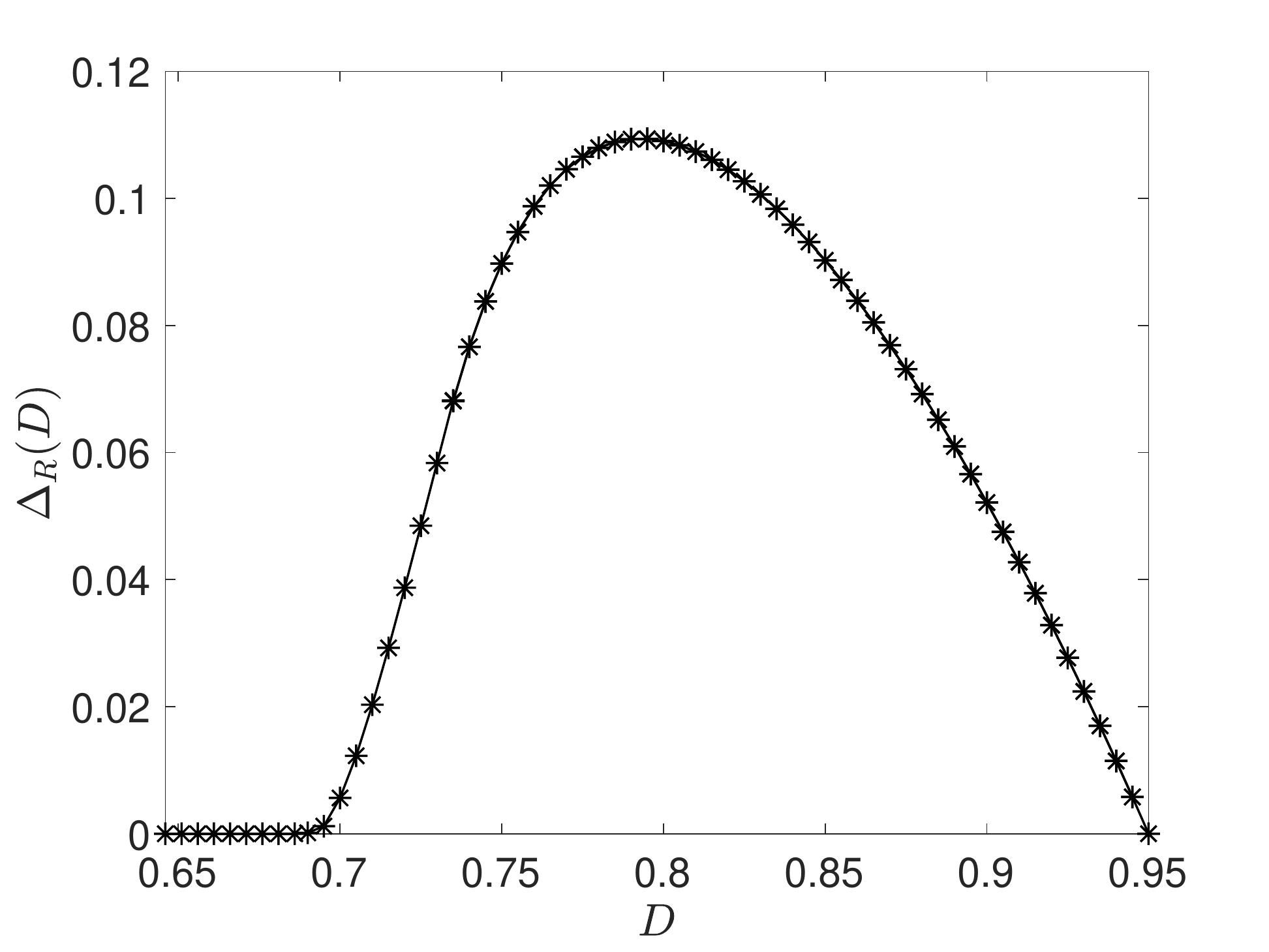}
\subcaption{}
\label{fig:delta_R_case2}
\end{minipage}\hfill
\begin{minipage}{0.5\linewidth}
\centering
\includegraphics[width=0.7\linewidth,height=4cm]{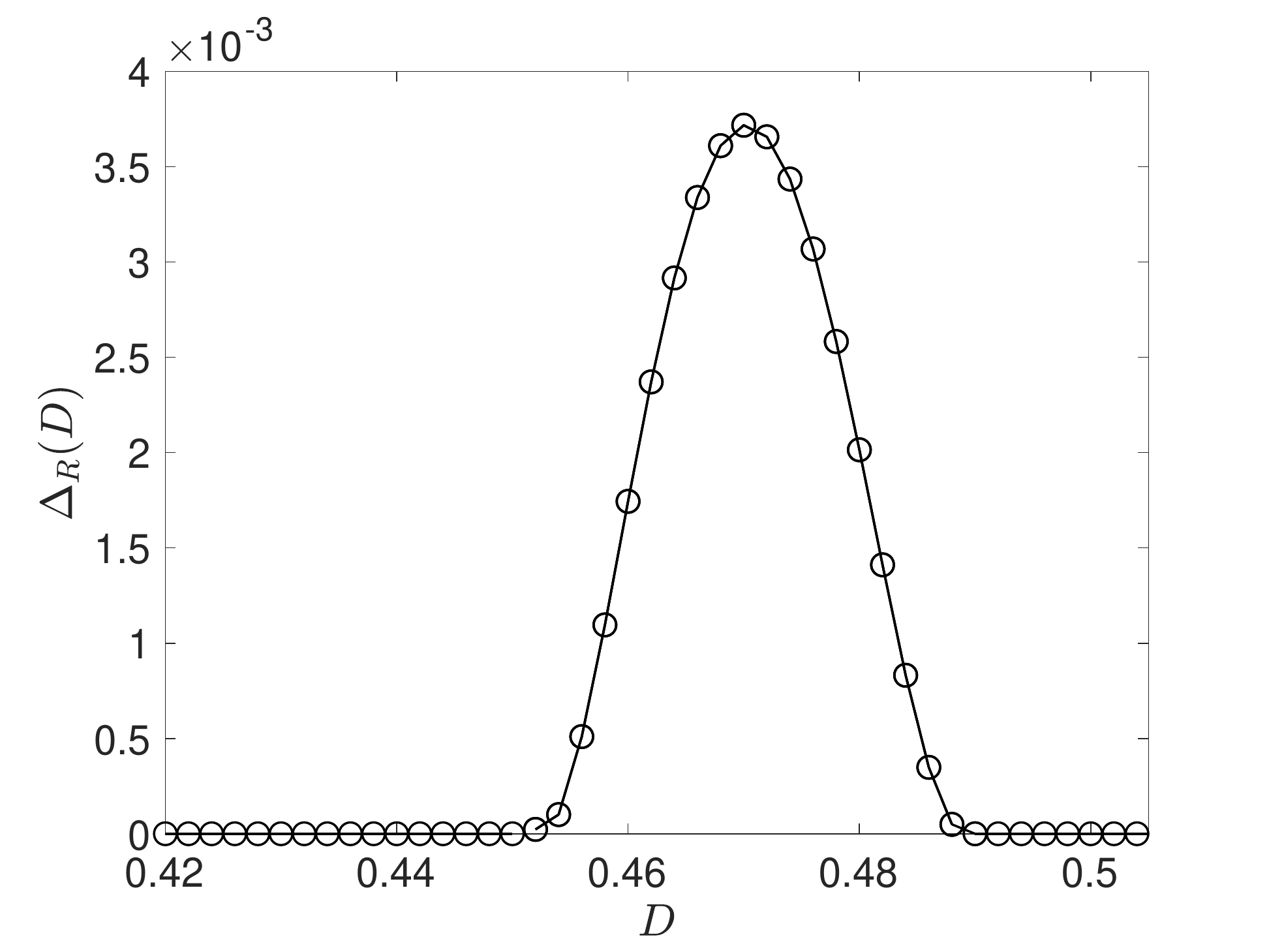}
\subcaption{}
\label{fig:delta_R_case3}
\end{minipage}
\caption{(a) Upper bound $\overline{\mathcal{R}}(D)$ with $D\in (d_{\min},\sigma_X^2)$ for the three cases. (b) Lower bound $\underline{\mathcal{R}}(D)$ with $D\in (d_{\min},\sigma_X^2)$ for the three cases. (c) $\Delta_{R}(D)$ with $D\in (d_{\min}, \sigma_X^2)$ for Case 2. (d) $\Delta_{R}(D)$ with $D\in (d_{\min},\sigma_X^2)$ for Case 3.}
\end{figure}

\underline{\textit{Example 1}:} In this example, we compare the upper bound $\overline{\mathcal{R}}(D)$ with the lower bound $\underline{\mathcal{R}}(D)$. We set $L=10$. In Fig.~\ref{fig:R_UB} and Fig. \ref{fig:R_LB}, we plot $\overline{\mathcal{R}}(D)$ and $\underline{\mathcal{R}}(D)$  with $D\in (d_{\min},\sigma_X^2)$ for the following three cases.
 \begin{itemize}
     \item Case 1: $\lambda_{X}=0.8$, $\gamma_{X}=1$, $\lambda_{Y}=5$, and $\gamma_{Y}=4$. In this case, we have $d_{\text{min}}=0.7422$ and $\sigma_X^2=0.98$. As can be seen from the figure, $\underline{\mathcal{R}}(D)$ coincides with $\overline{\mathcal{R}}(D)$ for all $D\in (d_{\min},\sigma_X^2)$, so $\mathcal{R}(D)$ is completely determined.
 \item Case 2: $\lambda_{X}=0.5$, $\gamma_{X}=1$, $\lambda_{Y}=6$, and $\gamma_{Y}=3$. In this case, we have $d_{\text{min}}\approx 0.646$, $\mathsf{D}_{\text{th},1}\approx 0.691$, $\mathsf{D}^c_{\text{th}}\approx0.733$ and $\sigma_X^2=0.95$. As  can be observed from both figures, 
 $\underline{\mathcal{R}}(D)$ coincides with $\overline{\mathcal{R}}(D)$ for $D \in (d_{\text{min}}, \mathsf{D}_{\text{th},1}]$ and consequently $\mathcal{R}(D)$ is determined over this interval (see the diamond-line portion of Fig. \ref{fig:R_LB}).   
 For $D \in (\mathsf{D}_{\text{th},1},\mathsf{D}^c_{\text{th}}]$, 
 $\underline{\mathcal{R}}(D)$ is characterized by $\underline{\mathcal{R}}_1^c(D)$  (see the plus-line portion of Fig. \ref{fig:R_LB}). For $D \in (\mathsf{D}^c_{\text{th}},\sigma_X^2)$,
  $\underline{\mathcal{R}}(D)$ is characterized by $\underline{\mathcal{R}}_2^c(D)$ (see the cross-line portion of Fig. \ref{fig:R_LB}). 
 
  \item Case 3: $\lambda_{X}=1$, $\gamma_{X}=0.45$, $\lambda_{Y}=12$, and $\gamma_{Y}=2.4$. In this case, we have $d_{\text{min}}=0.4207$, $\mathsf{D}_{\text{th},1}\approx 0.453$, $\mathsf{D}_{\text{th},2}\approx 0.489$ and $\sigma_X^2=0.505$.   For $D \in (d_{\text{min}}, \mathsf{D}_{\text{th},1}]$ and $D \in [\mathsf{D}_{\text{th},2},\sigma_X^2)$, 
 $\underline{\mathcal{R}}(D)$ coincides with $\overline{\mathcal{R}}(D)$ and consequently $\mathcal{R}(D)$ is determined over these two intervals (see the circle-line portion of Fig. \ref{fig:R_LB}). 
 For $D \in (\mathsf{D}_{\text{th},1},\mathsf{D}_{\text{th},2})$, 
 $\underline{\mathcal{R}}(D)$ is characterized by $\underline{\mathcal{R}}_1^c(D)$  (see the pentagonal-line portion of Fig. \ref{fig:R_LB}). 
 \end{itemize}
 
As can be observed from Fig.~\ref{fig:R_UB} and Fig.~\ref{fig:R_LB}, there exists a gap between  $\overline{\mathcal{R}}(D)$  and  $\underline{\mathcal{R}}(D)$ in Cases 2 and 3. We plot this gap, denoted by $\Delta_{R}(D)$, with $D\in (d_{\text{min}},\sigma^2_X)$ for these two cases in   Fig.~\ref{fig:delta_R_case2} and Fig.~\ref{fig:delta_R_case3}, respectively.

\setlength{\textfloatsep}{10pt plus 1.0pt minus 2.0pt}

	
		
    		

Now, we proceed to study the asymptotic behavior of the rate-distortion bounds $\overline{\mathcal{R}}(D)$ and $\underline{\mathcal{R}}(D)$ when $L$ tends to infinity. In the discussion below, it is necessary to assume that $\rho_X,\rho_Z\in [0,1]$. First, we perform the asymptotic analysis for $\overline{\mathcal{R}}(D)$. Define
\begin{IEEEeqnarray}{rCl}
d_{\min}^{\infty}&:=&\left\{\begin{array}{ll}\frac{\sigma_X^2\sigma_Z^2}{\sigma_X^2+\sigma_Z^2},&\hspace{1cm} \rho_X\sigma_X^2+\rho_Z\sigma_Z^2=0,\\\frac{\rho_X\rho_Z\sigma_X^2\sigma_Z^2}{\rho_X\sigma_X^2+\rho_Z\sigma_Z^2}+\gamma_X\gamma_Z\gamma_Y^{-1},&\hspace{1cm} \rho_X\sigma_X^2+\rho_Z\sigma_Z^2>0,\end{array}\right.\\
\mathsf{D}^{\infty}_{\text{th},0}&:=&\frac{\rho_X\rho_Z\sigma_X^2\sigma_Z^2}{\rho_X\sigma_X^2+\rho_Z\sigma_Z^2}+\gamma_X,\\
\xi &:=& \left(\frac{\rho_X}{1-\rho_X}\right)\cdot\left(\frac{1-\rho_Y}{\rho_Y}\right),
\end{IEEEeqnarray}
and 
\begin{IEEEeqnarray}{rCl}
\overline{\mathcal{R}}^{\infty}(D)&:=& \frac{L}{2}\log \frac{\sigma_X^4}{(\sigma_X^2+\sigma_Z^2)D-\sigma_X^2\sigma_Z^2},\\
\overline{\mathcal{R}}_1^{\infty}(D)&:=& \frac{L}{2}\log \frac{\gamma_X^2\gamma_Y^{-1}}{D-d^{\infty}_{\min}}+\frac{1}{2}\log L+\frac{1}{2}\log \frac{(\rho_X\sigma_X^2+\rho_Z\sigma_Z^2)(\mathsf{D}^{\infty}_{\text{th},0}-D)}{\gamma_X^2}\nonumber\\
&&\hspace{1cm}+\frac{ (\mathsf{D}^{\infty}_{\text{th},0}-\xi\gamma_X^2\gamma_Y^{-1}-D)^2}{2(\mathsf{D}^{\infty}_{\text{th},0}-D)(D-d^{\infty}_{\min}) }+O\left(\frac{1}{L}\right),\label{R1-exp}\\[1.5ex]
\overline{\mathcal{R}}_2^{\infty}(D) &:=& \frac{1}{2}\xi\sqrt{L}+\frac{1}{4}\log L+\frac{1}{2}\log(\frac{\rho_X}{1-\rho_X})+ \frac{\xi(\rho_X\gamma_X-\gamma_Z+(1-\rho_X^2)\sigma_Z^2)}{4(\rho_X^2\sigma_X^2+\rho_X\rho_Z\sigma_Z^2)}+O\left(\frac{1}{\sqrt{L}}\right),\nonumber\\\label{R2-upper-new}\\[1.5ex]
\overline{\mathcal{R}}_3^{\infty}(D) &:=& \frac{1}{2}\log \frac{\rho_X^2\sigma_X^4}{(\rho_X\sigma_X^2+\rho_Z\sigma_Z^2)(D-\mathsf{D}^{\infty}_{\text{th},0})}+\frac{(1-\rho_Y)(\sigma_X^2-D)}{2\rho_Y(D-\mathsf{D}^{\infty}_{\text{th},0})}+O\left(\frac{1}{L}\right).\label{R3-upper-new}
\end{IEEEeqnarray}
\begin{theorem}[Asymptotic Expression of Upper Bound]\label{upper-asym} 
\begin{enumerate}\item If $\rho_X\sigma_X^2+\rho_Z\sigma_Z^2=0$, then for $D\in (d_{\min}^{\infty},\sigma_X^2)$, we have
\begin{IEEEeqnarray}{rCl}
\overline{\mathcal{R}}(D)= \overline{\mathcal{R}}^{\infty}(D).\label{approx-upp1}
\end{IEEEeqnarray}
\item If $\rho_X\sigma_X^2+\rho_Z\sigma_Z^2>0$, then for $D\in (d_{\min}^{\infty},\sigma_X^2)$, we have
\begin{IEEEeqnarray}{rCl}
\overline{\mathcal{R}}(D)= \left\{\begin{array}{ll} \overline{\mathcal{R}}_1^{\infty}(D), & D<\mathsf{D}^{\infty}_{\text{th},0},\\ \overline{\mathcal{R}}_2^{\infty}(D), & D=\mathsf{D}^{\infty}_{\text{th},0},\\ \overline{\mathcal{R}}_3^{\infty}(D), & D>\mathsf{D}^{\infty}_{\text{th},0}.\end{array} \right.\label{approx-upp2}
\end{IEEEeqnarray}
\end{enumerate}
\end{theorem}
\begin{IEEEproof} See Section \ref{upper-asym-proof}.
\end{IEEEproof}

Next, we perform the asymptotic analysis for $\underline{\mathcal{R}}(D)$. Define
\begin{IEEEeqnarray}{rCl}
\mathsf{D}^{\infty}_{\text{th},1}&:=& \frac{\rho_X\rho_Z\sigma_X^2\sigma_Z^2}{\rho_X\sigma_X^2+\rho_Z\sigma_Z^2}+\gamma_X-\frac{1+\sqrt{1-4\xi^2}}{2}\gamma_X^2\gamma_Y^{-1},\\
\mathsf{D}^{\infty}_{\text{th},2}&:=& \frac{\rho_X\rho_Z\sigma_X^2\sigma_Z^2}{\rho_X\sigma_X^2+\rho_Z\sigma_Z^2}+\gamma_X-\frac{1-\sqrt{1-4\xi^2}}{2}\gamma_X^2\gamma_Y^{-1},\\
\end{IEEEeqnarray}
and 
\begin{IEEEeqnarray}{rCl}
\underline{\mathcal{R}}_1^{\infty}(D)&:=&\frac{L+1}{2}\log \frac{\gamma_X^2\gamma_Y^{-1}}{D-d^{\infty}_{\min}}+\frac{1}{2}\log L+\frac{1}{2}\frac{(1-2\xi)\gamma_X^2\gamma_Y^{-1}}{D-d^{\infty}_{\min}}+\frac{1}{2}\log \frac{\rho_X^2(1-\rho_Y)}{(1-\rho_X)^2\rho_Y}+O\left(\frac{1}{L}\right),\nonumber\\\label{R1-lower-new}\\
\underline{\mathcal{R}}_2^{\infty}(D)&:=&\frac{L}{2}\log \frac{\sigma_X^4}{\gamma_YD-\sigma_X^2\gamma_Z}-\frac{1}{2}\frac{D-\sigma_X^2}{D-\sigma_X^2+\sigma_X^4\gamma_Y^{-1}}+O\left(\frac{1}{L}\right).\\
\end{IEEEeqnarray}
\begin{theorem}[Asymptotic Expression of Lower Bound]\label{lower-asym-thm} 
\begin{enumerate}
    \item If $\rho_X\sigma_X^2+\rho_Z\sigma_Z^2=0$, then for $D\in (d_{\min}^{\infty},\sigma_X^2)$, we have
    \begin{IEEEeqnarray}{rCl}
\underline{\mathcal{R}}(D)= \overline{\mathcal{R}}^{\infty}(D).\label{lower-thm-cond1}
\end{IEEEeqnarray}

\item If $\rho_X\sigma_X^2+\rho_Z\sigma_Z^2>0$, $\rho_X>0$ and $\xi\geq \frac{1}{2}$,  then for $D\in (d_{\min}^{\infty},\sigma_X^2)$, we have
\begin{IEEEeqnarray}{rCl}
&&\hspace{-0.5cm}\underline{\mathcal{R}}(D)= \left\{\begin{array}{ll}
\overline{\mathcal{R}}_1^{\infty}(D), &  D< \mathsf{D}^{\infty}_{\text{th},0},\\ \overline{\mathcal{R}}_2^{\infty}(D), & D=\mathsf{D}^{\infty}_{\text{th},0}, \\ \overline{\mathcal{R}}_3^{\infty}(D), & D>\mathsf{D}^{\infty}_{\text{th},0}.\end{array} \right.\label{rate-final-lower-2nd-condition}
\end{IEEEeqnarray}\normalsize\vspace{0cm}
\item If $\rho_X\sigma_X^2+\rho_Z\sigma_Z^2>0$, $\rho_X>0$ and $\xi< \frac{1}{2}$,  then for $D\in (d_{\min}^{\infty},\sigma_X^2)$, we have
\begin{IEEEeqnarray}{rCl}
&&\hspace{-0.5cm}\underline{\mathcal{R}}(D)= \left\{\begin{array}{ll}\overline{\mathcal{R}}^{\infty}_1(D), & D\leq \mathsf{D}^{\infty}_{\text{th},1},\\[0.8ex]
\underline{\mathcal{R}}_1^{\infty}(D), &  \mathsf{D}^{\infty}_{\text{th},1}<  D<  \mathsf{D}^{\infty}_{\text{th},2},\\[0.5ex]
\overline{\mathcal{R}}_1^{\infty}(D), & \mathsf{D}^{\infty}_{\text{th},2}\leq D< \mathsf{D}^{\infty}_{\text{th},0}, \\[0.5ex]
 \overline{\mathcal{R}}_2^{\infty}(D), & D=\mathsf{D}^{\infty}_{\text{th},0},\\ \overline{\mathcal{R}}_3^{\infty}(D), & D>\mathsf{D}^{\infty}_{\text{th},0}.\end{array} \right.\label{rate-final-lower-3rd-condition}
\end{IEEEeqnarray}
\item If $\rho_X\sigma_X^2+\rho_Z\sigma_Z^2>0$ and $\rho_X=0$,  then for $D\in (d_{\min}^{\infty},\sigma_X^2)$, we have
\begin{IEEEeqnarray}{rCl}
\underline{\mathcal{R}}(D)=\underline{\mathcal{R}}^{\infty}_2(D).\label{rate-final-lower-4th-condition}
\end{IEEEeqnarray}
\end{enumerate}

\end{theorem}
\begin{IEEEproof} See Section \ref{lower-asym-thm-proof}. 
\end{IEEEproof}

The following corollary provides the (asymptotic) gap between $\overline{\mathcal{R}}(D)$ and $\underline{\mathcal{R}}(D)$. Define
\begin{IEEEeqnarray}{rCl}
&&\Delta^{(\infty)}_R(D):= 
\hspace{0.1cm}\frac{(\mathsf{D}^{\infty}_{\text{th},1}-D)(\mathsf{D}^{\infty}_{\text{th},2}-D)}{2\left(\mathsf{D}^{\infty}_{\text{th},0}-D\right)\left(D-d^{\infty}_{\min}\right)}+\frac{1}{2}\log \frac{\gamma_Y^2}{\xi^2\gamma_X^4} \left(\mathsf{D}^{\infty}_{\text{th},0}-D\right)\left(D-d^{\infty}_{\min}\right).
\end{IEEEeqnarray}

\begin{figure}
    \centering
    \includegraphics[scale=0.35]{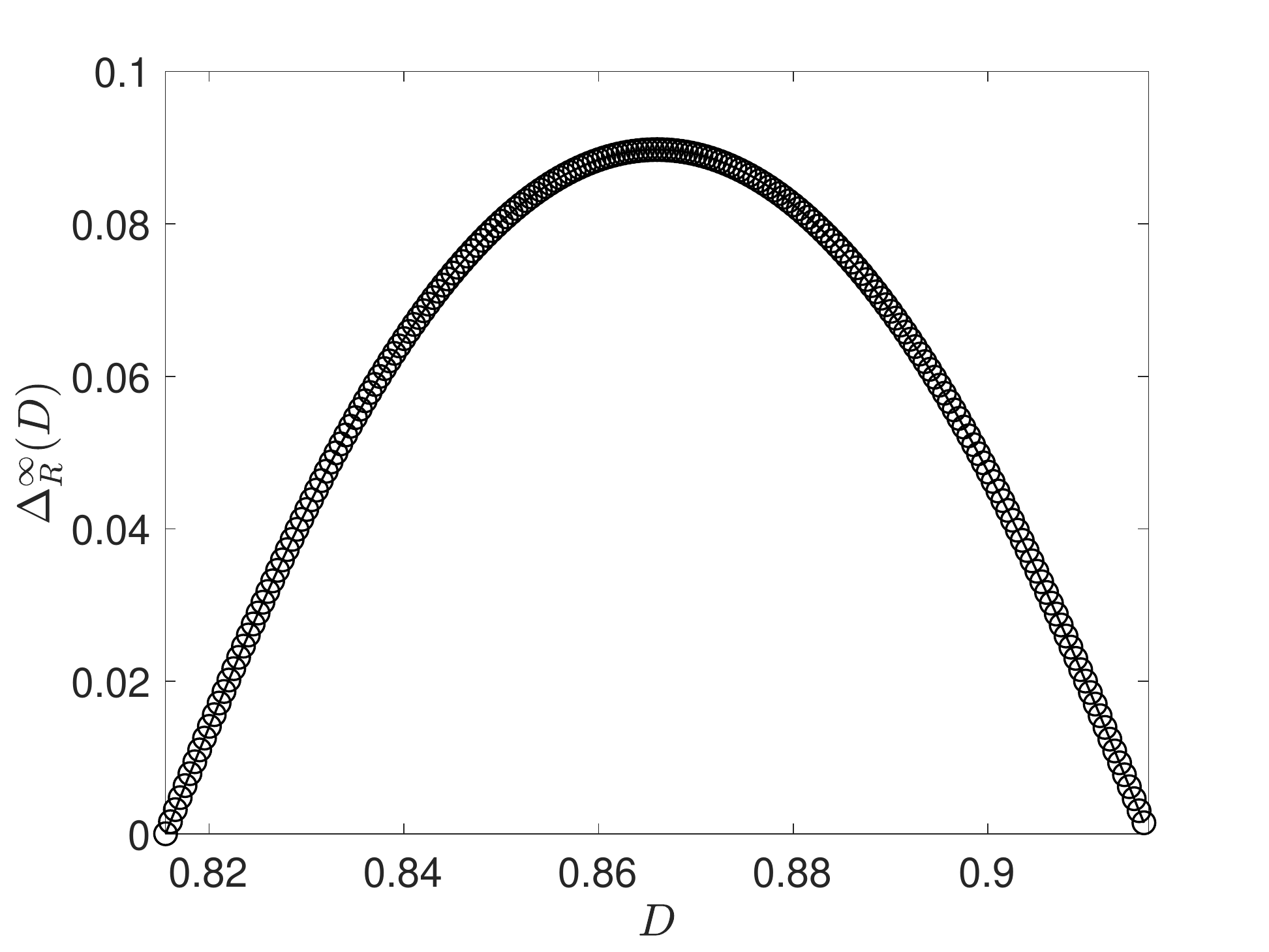}
    \caption{$\Delta^{\infty}_R(D)$ with $\rho_X=0.3$, $\sigma_X^2=1$, $\rho_Y=0.5$, $\sigma_Y^2=5$ and $D \in(\mathsf{D}^{\infty}_{\text{th},1},\mathsf{D}^{\infty}_{\text{th},2})$, where $\mathsf{D}^{\infty}_{\text{th},1}\approx0.816$ and $\mathsf{D}^{\infty}_{\text{th},2}\approx 0.917$.}
    \label{gap-asymptotic}
\end{figure}

\begin{corollary} The  gap between $\overline{\mathcal{R}}(D)$ and $\underline{\mathcal{R}}(D)$ is given as follows.
\begin{enumerate}
    \item If $\rho_X\sigma_X^2+\rho_Z\sigma_Z^2=0$, then for $D\in (d_{\min}^{\infty},\sigma_X^2)$, we have
    \begin{IEEEeqnarray}{rCl}
\overline{\mathcal{R}}(D)- \underline{\mathcal{R}}(D)=0.
\end{IEEEeqnarray}
\item If $\rho_X\sigma_X^2+\rho_Z\sigma_Z^2>0$, $\rho_X>0$ and $\xi\geq \frac{1}{2}$, then for $D\in (d_{\min}^{\infty},\sigma_X^2)$, we have
\begin{IEEEeqnarray}{rCl}
\overline{\mathcal{R}}(D)- \underline{\mathcal{R}}(D)=0.
\end{IEEEeqnarray}
\item If $\rho_X\sigma_X^2+\rho_Z\sigma_Z^2>0$, $\rho_X>0$ and $\xi<\frac{1}{2}$, then for $D\in (d_{\min}^{\infty},\sigma_X^2)$, we have
\begin{IEEEeqnarray}{rCl}
\lim_{L\to\infty}\overline{\mathcal{R}}(D)- \underline{\mathcal{R}}(D)=\left\{\begin{array}{ll}0,& D\leq \mathsf{D}_{\text{th},1}^{\infty}\;\;\;\text{or}\;\;\;D\geq\mathsf{D}^{\infty}_{\text{th},2},\\ \Delta^{\infty}_R(D), & \mathsf{D}^{\infty}_{\text{th},1}<D<\mathsf{D}^{\infty}_{\text{th},2}.\end{array}\right.
\end{IEEEeqnarray}

\item If $\rho_X\sigma_X^2+\rho_Z\sigma_Z^2>0$ and $\rho_X=0$, then for $D\in (d_{\min}^{\infty},\sigma_X^2)$, we have
\begin{IEEEeqnarray}{rCl}
\overline{\mathcal{R}}(D)- \underline{\mathcal{R}}(D)=O(\log L).
\end{IEEEeqnarray}
\end{enumerate}
\end{corollary}
As  can be seen from the above corollary, under the third condition, the lower and upper bounds asymptotically match  for all $D$ except when $\mathsf{D}^{\infty}_{\text{th},1}<D<\mathsf{D}^{\infty}_{\text{th},2}$. Fig.~\ref{gap-asymptotic} plots the function $\Delta^{(\infty)}_R(D)$, which characterizes the asymptotic gap between $\overline{\mathcal{R}}(D)$ and $\underline{\mathcal{R}}(D)$ (as $L$ tends to infinity) 
in the interval $\mathsf{D}^{\infty}_{\text{th},1}<D<\mathsf{D}^{\infty}_{\text{th},2}$, for some values of parameters.


\section{Proof of Theorem~\ref{conv-new}}\label{conv-new-proof}
Before starting the proof, we introduce another representation of $\overline{\mathcal{R}}(D)$ (defined in \eqref{ach-rate}--\eqref{distortion-ach}) which will be repeatedly used in the sequel. Define 
\begin{IEEEeqnarray}{rCl}
\lambda_I^{-1}&:=&\lambda_Y^{-1}+\lambda_Q^{-1},\\
\gamma_I^{-1}&:=&\gamma_Y^{-1}+\lambda_Q^{-1}.
\end{IEEEeqnarray}
\begin{corollary}\label{cor-new-RD} $\overline{\mathcal{R}}(D)$ can be alternatively expressed as
\begin{IEEEeqnarray}{rCl}
\overline{\mathcal{R}}(D)=\frac{1}{2}\log (\lambda_I^{-1}\lambda_Y)+\frac{L-1}{2}\log \left(1+\gamma_Y(\lambda_I^{-1}-\lambda_Y^{-1})\right),\label{RD-new-form}
\end{IEEEeqnarray}
where
\begin{IEEEeqnarray}{rCl}
\lambda_X^2\lambda_Y^{-2}\lambda_I+\lambda_X-\lambda_X^2\lambda_Y^{-1}+(L-1)(\gamma_X^2\gamma_Y^{-2}(\lambda_I^{-1}+\gamma_Y^{-1}-\lambda_Y^{-1})^{-1}+\gamma_X-\gamma_X^2\gamma_Y^{-1})=LD,\nonumber\\\label{distortion-new-form}
\end{IEEEeqnarray}
or in the following form
\begin{IEEEeqnarray}{rCl}
\overline{\mathcal{R}}(D)=\frac{1}{2}\log\left(1+\lambda_Y(\gamma_I^{-1}-\gamma_Y^{-1})\right)+\frac{L-1}{2}\log(\gamma_I^{-1}\gamma_Y),
\end{IEEEeqnarray}
where 
\begin{IEEEeqnarray}{rCl}
\lambda_X^2\lambda_Y^{-2}(\gamma_{I}^{-1}+\lambda_Y^{-1}-\gamma_Y^{-1})^{-1}+\lambda_X-\lambda_X^2\lambda_Y^{-1}+(L-1)(\gamma_X^2\gamma_Y^{-2}\gamma_I+\gamma_X-\gamma_X^2\gamma_Y^{-1})=LD.
\end{IEEEeqnarray}
\end{corollary}
Now, consider the optimization problem in Theorem~\ref{thm-conv} as follows:
\begin{IEEEeqnarray}{rCl}
&&\hspace{1cm}\min_{\alpha,\beta,\delta}\Omega(\alpha,\beta,\delta),\\
&&\text{s.t.} \;\;\;\; \text{constraints}\;\; \eqref{opt-consa}-\eqref{opt-consz}.
\end{IEEEeqnarray}
Based on the fact that $\lambda_Y\geq \gamma_Y$ or $\gamma_Y\geq \lambda_Y$, we get two different cases.

First, consider the case $\lambda_Y\geq \gamma_Y> 0$, where we have $\lambda_W=\gamma_Y$. Thus, the objective function reduces to 
\begin{IEEEeqnarray}{rCl}
&&\hspace{1cm}f(\alpha,\delta):= \frac{1}{2}\log \frac{\lambda_Y^2}{(\lambda_Y-\gamma_Y)\alpha+\lambda_Y\gamma_Y}+\frac{L}{2}\log \frac{\gamma_Y}{\delta},\label{obj-function}
\end{IEEEeqnarray}
and the constraints \eqref{opt-consa}-\eqref{opt-consz} are simplified as follows:
\begin{subequations}
\begin{IEEEeqnarray}{rCl}
&& 0<\alpha\leq \lambda_Y,\\
&& 0<\beta\leq \gamma_Y,\label{opt2-cons1}\\
&& 0<\delta,\\
&& \delta \leq (\alpha^{-1}+\gamma_Y^{-1}-\lambda_Y^{-1})^{-1},\\
&& \delta \leq \beta,\label{opt2-cons2}\\
&& \lambda_X^2\lambda_Y^{-2}\alpha+\lambda_X-\lambda_X^2\lambda_Y^{-1}+(L-1)(\gamma_X^2\gamma_Y^{-2}\beta+\gamma_X-\gamma_X^2\gamma_Y^{-1})\leq LD.\label{opt2-cons3}
\end{IEEEeqnarray}
\end{subequations}
Since the objective function does not depend on parameter $\beta$, we can eliminate $\beta$ from the constraints \eqref{opt2-cons1}, \eqref{opt2-cons2} and \eqref{opt2-cons3}. Thus, we get the following new constraints:
\begin{subequations}\begin{IEEEeqnarray}{rCl}
&& 0<\alpha\leq \lambda_Y,\label{opt3-cons1}\\
&&0<\delta,\\
&& \delta \leq (\alpha^{-1}+\gamma_Y^{-1}-\lambda_Y^{-1})^{-1},\label{opt3-cons2}\\
&&\delta\leq \gamma_Y,\label{opt3-cons3}\\
&& \lambda_X^2\lambda_Y^{-2}\alpha+\lambda_X-\lambda_X^2\lambda_Y^{-1}+(L-1)(\gamma_X^2\gamma_Y^{-2}\delta+\gamma_X-\gamma_X^2\gamma_Y^{-1})\leq LD.
\end{IEEEeqnarray}\end{subequations}
Given constraint \eqref{opt3-cons1}, the inequality \eqref{opt3-cons2} is more restricting compared to \eqref{opt3-cons3}, so the above constraints reduce to  
\begin{subequations}\label{opt4-case1}
\begin{IEEEeqnarray}{rCl}
&&0<\alpha\leq \lambda_Y,\\
&&0<\delta \leq  (\alpha^{-1}+\gamma_Y^{-1}-\lambda_Y^{-1})^{-1},\\
&&\lambda_X^2\lambda_Y^{-2}\alpha+\lambda_X-\lambda_X^2\lambda_Y^{-1}+(L-1)(\gamma_X^2\gamma_Y^{-2}\delta+\gamma_X-\gamma_X^2\gamma_Y^{-1})\leq LD.
\end{IEEEeqnarray}
 \end{subequations}
 Then, the goal is to minimize $f(\alpha,\delta)$ subject to the constraints \eqref{opt4-case1}, which is a convex program. According to the KKT optimality conditions, there exist nonnegative Lagrange multipliers $\{\omega_1,\omega_2,\omega_3\}$ and optimal solutions $(\alpha^*,\delta^*)$ such that
 \begin{subequations}\label{kkt-conditions}
 \begin{IEEEeqnarray}{rCl}
 &&\frac{\gamma_Y-\lambda_Y}{2((\lambda_Y-\gamma_Y)\alpha^*+\lambda_Y\gamma_Y)}+\omega_1-\omega_2(1+(\gamma_Y^{-1}-\lambda_Y^{-1})\alpha^*)^{-2}+\omega_3\lambda_X^2\lambda_Y^{-2}=0,\\
 &&-\frac{L}{2\delta^*}+\omega_2+(L-1)\omega_3\gamma_X^2\gamma_Y^{-2}=0,\\
 &&\omega_1(\alpha^*-\lambda_Y)=0,\\
 &&\omega_2(\delta^*-((\alpha^*)^{-1}+\gamma_Y^{-1}-\lambda_Y^{-1})^{-1})=0,\label{o2-cond}\\
 &&\omega_3(\lambda_X^2\lambda_Y^{-2}\alpha^*+\lambda_X-\lambda_X^2\lambda_Y^{-1}+(L-1)(\gamma_X^2\gamma_Y^{-2}\delta^*+\gamma_X-\gamma_X^2\gamma_Y^{-1})-LD)=0.
 \end{IEEEeqnarray}
\end{subequations}

In the following, we consider two different cases for the Lagrange multipliers.

 \underline{\textbf{Case 1 ($\omega_2>0$):}}
    
    In this case, the KKT conditions in \eqref{kkt-conditions} reduce to
    \begin{subequations}\label{kkt-con2}
    \begin{IEEEeqnarray}{rCl}
    &&\frac{\gamma_Y-\lambda_Y}{2((\lambda_Y-\gamma_Y)\alpha^*+\lambda_Y\gamma_Y)}+\omega_1-\omega_2(1+(\gamma_Y^{-1}-\lambda_Y^{-1})\alpha^*)^{-2}+\omega_3\lambda_X^2\lambda_Y^{-2}=0,\\
 &&-\frac{L}{2\delta^*}+\omega_2+(L-1)\omega_3\gamma_X^2\gamma_Y^{-2}=0,\\
 &&\omega_1(\alpha^*-\lambda_Y)=0,\\
 &&\delta^*-((\alpha^*)^{-1}+\gamma_Y^{-1}-\lambda_Y^{-1})^{-1}=0,\label{aldel}\\
 &&\omega_3(\lambda_X^2\lambda_Y^{-2}\alpha^*+\lambda_X-\lambda_X^2\lambda_Y^{-1}+(L-1)(\gamma_X^2\gamma_Y^{-2}\delta^*+\gamma_X-\gamma_X^2\gamma_Y^{-1})-LD)=0.
    \end{IEEEeqnarray}
    \end{subequations}
    Assume that $\alpha^*$ and $\delta^*$ satisfy 
    \begin{IEEEeqnarray}{rCl}
    &&\lambda_X^2\lambda_Y^{-2}\alpha^*+\lambda_X-\lambda_X^2\lambda_Y^{-1}+(L-1)(\gamma_X^2\gamma_Y^{-2}\delta^*+\gamma_X-\gamma_X^2\gamma_Y^{-1})=LD.\label{distortion-cons}
    \end{IEEEeqnarray}
    Solving the set of equations in \eqref{kkt-con2} yields
    \begin{subequations}
    \begin{IEEEeqnarray}{rCl}
    &&\omega_1=0,\\
    &&\omega_2=\frac{L}{2\delta^*}-(L-1)\omega_3\gamma_X^2\gamma_Y^{-2},\\
    &&\omega_3 =\frac{\frac{L}{2\delta^*}(1+(\gamma_Y^{-1}-\lambda_Y^{-1})\alpha^*)^{-2}+\frac{1}{2}(\gamma_Y^{-1}-\lambda_Y^{-1})(1+(\gamma_Y^{-1}-\lambda_Y^{-1})\alpha^*)^{-1}}{\lambda_X^2\lambda_Y^{-2}+(L-1)\gamma_X^2\gamma_Y^{-2}(1+(\gamma_Y^{-1}-\lambda_Y^{-1})\alpha^*)^{-2}}.
    \end{IEEEeqnarray}
    \end{subequations}
    Notice that $\omega_3\geq 0$ since $\lambda_Y\geq \gamma_Y$. We should make sure that $\omega_2 \geq 0$. This gives the following inequality:
    \begin{IEEEeqnarray}{rCl}
    \frac{1}{2L}(\gamma_Y^{-1}-\lambda_Y^{-1})(1+(\gamma_Y^{-1}-\lambda_Y^{-1})\alpha^*)^{-1}\leq \frac{1}{2(L-1)\delta^*}\lambda_X^2\gamma_X^{-2}\lambda_Y^{-2}\gamma_Y^{2},
    \end{IEEEeqnarray}
    which can be equivalently written as 
    \begin{IEEEeqnarray}{rCl}
     \delta^*\leq \frac{L}{L-1}\lambda_X^2\gamma_X^{-2}\lambda_Y^{-2}\gamma_Y^2\left(\left(\gamma_Y^{-1}-\lambda_Y^{-1}\right)^{-1}+\alpha^*\right).\label{delta-alpha-constraint}
    \end{IEEEeqnarray}
    Combining the above inequality with \eqref{distortion-cons}, we can write
    \begin{IEEEeqnarray}{rCl}
    &&LD\leq L\lambda_X^2(\lambda_Y-\gamma_Y)^{-1}+(L-1)\gamma_X^2(\gamma_X^{-1}-\gamma_Y^{-1})+\lambda_X-(L+1)\lambda_X^2\lambda_Y^{-1}+(L+1)\alpha^*\lambda_X^2\lambda_Y^{-2}.\nonumber\\\label{D-ineq-proof}
    \end{IEEEeqnarray}
    Define
    \begin{IEEEeqnarray}{rCl}\lambda_I:=\alpha^*.\label{lI-al-def}\end{IEEEeqnarray}
    Considering \eqref{delta-alpha-constraint} with \eqref{aldel} and re-arranging the terms yields the following constraint:
    \begin{IEEEeqnarray}{rCl}
    (L-1)\gamma_X^2\gamma_Y^{-2}(\lambda_I^{-1}+\gamma_Y^{-1}-\lambda_Y^{-1})^{-1}-L\lambda_I\lambda_X^2\lambda_Y^{-2}&\leq& L\lambda_X^2\lambda_Y^{-1}(\lambda_Y\gamma_Y^{-1}-1)^{-1}.\label{condition1-ineq}
    \end{IEEEeqnarray}
     Re-arranging the terms in \eqref{D-ineq-proof} and \eqref{distortion-cons}, we have 
    \begin{IEEEeqnarray}{rCl}
&&LD\leq L\lambda_X^2\lambda_Y^{-1}(\lambda_Y\gamma_Y^{-1}-1)^{-1}+(L-1)\gamma_X^2(\gamma_X^{-1}-\gamma_Y^{-1})+\lambda_X-\lambda_X^2\lambda_Y^{-1}+(L+1)\lambda_I\lambda_X^2\lambda_Y^{-2},\nonumber\\
&&LD=\lambda_X-\lambda_X^2\lambda_Y^{-1}+(L-1)(\gamma_X^2\gamma_Y^{-2}(\lambda_I^{-1}+\gamma_Y^{-1}-\lambda_Y^{-1})^{-1}+\gamma_X-\gamma_X^2\gamma_Y^{-1})+\lambda_X^2\lambda_Y^{-2}\lambda_I.
\end{IEEEeqnarray}
Thus, we define the following set as the admissible distortion set:
\begin{IEEEeqnarray}{rCl}
&&\mathcal{D}_1(\lambda_I):=\{D\in (d_{\min},\sigma_X^2)\colon \nonumber\\&&\hspace{0.2cm}LD\leq L\lambda_X^2\lambda_Y^{-1}(\lambda_Y\gamma_Y^{-1}-1)^{-1}+(L-1)\gamma_X^2(\gamma_X^{-1}-\gamma_Y^{-1})+\lambda_X-\lambda_X^2\lambda_Y^{-1}+(L+1)\lambda_I\lambda_X^2\lambda_Y^{-2},\nonumber\\
&&\hspace{0.2cm}LD=\lambda_X-\lambda_X^2\lambda_Y^{-1}+(L-1)(\gamma_X^2\gamma_Y^{-2}(\lambda_I^{-1}+\gamma_Y^{-1}-\lambda_Y^{-1})^{-1}+\gamma_X-\gamma_X^2\gamma_Y^{-1})+\lambda_X^2\lambda_Y^{-2}\lambda_I\}.\label{D1set}
\end{IEEEeqnarray}
 Plugging \eqref{aldel} into \eqref{obj-function} and considering \eqref{distortion-cons} yields the rate-distortion expression $\overline{\mathcal{R}}(D)$ defined in \eqref{RD-new-form} subject to constraint \eqref{distortion-new-form}. 
    \vspace{0.3cm}
    
 \underline{\textbf{Case 2 ($\omega_2=0$):}}
    
    In this case, the KKT conditions in \eqref{kkt-conditions} reduce to 
    \begin{subequations}\label{kkt-cond3}
    \begin{IEEEeqnarray}{rCl}
    &&\omega_1 = \frac{\lambda_Y-\gamma_Y}{2((\lambda_Y-\gamma_Y)\alpha^*+\lambda_Y\gamma_Y)}-\omega_3\lambda_X^2\lambda_Y^{-2},\\
     && \omega_3 = \frac{L}{2\delta^*(L-1)}\gamma_Y^2\gamma_X^{-2},\\
 &&\omega_1(\alpha^*-\lambda_Y)=0,\\
 &&\lambda_X^2\lambda_Y^{-2}\alpha^*+\lambda_X-\lambda_X^2\lambda_Y^{-1}+(L-1)(\gamma_X^2\gamma_Y^{-2}\delta^*+\gamma_X-\gamma_X^2\gamma_Y^{-1})-LD=0.
    \end{IEEEeqnarray}
    \end{subequations}
    To solve the above set of equations, we consider two different subcases: $\omega_1>0$ and $\omega_1 =0$.

 \underline{\textit{Subcase a ($\omega_1=0$):}}
    
    Solving the set of equations in \eqref{kkt-cond3} with $\omega_1=0$ yields
    \begin{subequations}
    \begin{IEEEeqnarray}{rCl}
    && \alpha^* = \frac{1}{2\omega_3}\lambda_X^{-2}\lambda_Y^2-(\gamma_Y^{-1}-\lambda_Y^{-1})^{-1},\label{alp-sol}\\
    &&\delta^* = \frac{L}{2(L-1)\omega_3}\gamma_Y^2\gamma_X^{-2},\label{del-sol}\\
    &&\omega_3=\frac{L+1}{2}\cdot \frac{1}{LD-\lambda_X-(L-1)(\gamma_X-\gamma_X^2\gamma_Y^{-1})+\lambda_X^2\lambda_Y^{-2}(\lambda_Y+(\gamma_Y^{-1}-\lambda_Y^{-1})^{-1})}.\nonumber\\\label{D-sol}
    \end{IEEEeqnarray} 
    \end{subequations}
 Recalling the definition of $\lambda_I$ in \eqref{lI-al-def}, considering \eqref{alp-sol} with \eqref{D-sol} and re-arranging the terms, we get the following equation:
 \begin{IEEEeqnarray}{rCl}
 LD&=& L\lambda_X^2\lambda_Y^{-1}(\lambda_Y\gamma_Y^{-1}-1)^{-1}+(L-1)\gamma_X^2(\gamma_X^{-1}-\gamma_Y^{-1})+\lambda_X-\lambda_X^2\lambda_Y^{-1}+(L+1)\lambda_I\lambda_X^2\lambda_Y^{-2}.\nonumber\\\label{Dc-cons}
 \end{IEEEeqnarray}
 Notice that \eqref{o2-cond} with $\omega_2=0$ implies that 
 \begin{IEEEeqnarray}{rCl}
 \delta^*<((\alpha^*)^{-1}+\gamma_Y^{-1}-\lambda_Y^{-1})^{-1}. \label{al-del-strict}
 \end{IEEEeqnarray}
 Moreover, \eqref{alp-sol} with the fact that $ \alpha^*< \lambda_Y$ gives 
 \begin{IEEEeqnarray}{rCl}
\omega_3> \frac{1}{2}\lambda_X^{-2}(\lambda_Y-\gamma_Y),
 \end{IEEEeqnarray}
 which together with \eqref{D-sol} yields the following constraint on $D$:
 \begin{IEEEeqnarray}{rCl}
LD< L\lambda_X^2(\lambda_Y-\gamma_Y)^{-1}+(L-1)\gamma_X^2(\gamma_X^{-1}-\gamma_Y^{-1})+\lambda_X.\label{subcase1-condition}
\end{IEEEeqnarray}
 Plugging \eqref{alp-sol} and \eqref{del-sol} into \eqref{al-del-strict} and re-arranging the terms give the following condition:
 \begin{IEEEeqnarray}{rCl}
 L\lambda_X^2\lambda_Y^{-1}(\lambda_Y\gamma_Y^{-1}-1)^{-1}&<& (L-1)\gamma_X^2\gamma_Y^{-2}(\lambda_I^{-1}+\gamma_Y^{-1}-\lambda_Y^{-1})^{-1}-L\lambda_I\lambda_X^2\lambda_Y^{-2}.\label{2nd-condition}
 \end{IEEEeqnarray}
 Combining \eqref{2nd-condition} with \eqref{Dc-cons} yields 
 \begin{IEEEeqnarray}{rCl}
 &&LD<\lambda_X-\lambda_X^2\lambda_Y^{-1}+(L-1)(\gamma_X^2\gamma_Y^{-2}(\lambda_I^{-1}+\gamma_Y^{-1}-\lambda_Y^{-1})^{-1}+\gamma_X-\gamma_X^2\gamma_Y^{-1})+\lambda_X^2\lambda_Y^{-2}\lambda_I.\;\label{Dc-cons2}
 \end{IEEEeqnarray}
 The conditions \eqref{Dc-cons} and \eqref{Dc-cons2} define the following distortion set:
 \begin{IEEEeqnarray}{rCl}
&&\mathcal{D}_1^c(\lambda_I):=\{D\in (d_{\min},\sigma_X^2)\colon \nonumber\\&&\hspace{0.2cm}LD= L\lambda_X^2\lambda_Y^{-1}(\lambda_Y\gamma_Y^{-1}-1)^{-1}+(L-1)\gamma_X^2(\gamma_X^{-1}-\gamma_Y^{-1})+\lambda_X-\lambda_X^2\lambda_Y^{-1}+(L+1)\lambda_I\lambda_X^2\lambda_Y^{-2},\nonumber\\
&&\hspace{0.2cm}LD<\lambda_X-\lambda_X^2\lambda_Y^{-1}+(L-1)(\gamma_X^2\gamma_Y^{-2}(\lambda_I^{-1}+\gamma_Y^{-1}-\lambda_Y^{-1})^{-1}+\gamma_X-\gamma_X^2\gamma_Y^{-1})+\lambda_X^2\lambda_Y^{-2}\lambda_I\}.\label{D1set-complement}
\end{IEEEeqnarray}
In summary, for this subcase, $D\in \mathcal{D}_1^c(\lambda_I)$ while the constraint \eqref{subcase1-condition} holds.
 Plugging \eqref{alp-sol}--\eqref{D-sol} into \eqref{obj-function} gives the rate-distortion expression $\underline{\mathcal{R}}_1^c(D)$ defined in~\eqref{R2Dl}.
 
      \underline{\textit{Subcase b ($\omega_1>0$):}}
    
    Here, we get the following solution to \eqref{kkt-cond3}:
    \begin{subequations}
    \begin{IEEEeqnarray}{rCl}
    &&\alpha^*=\lambda_Y,\label{case3-cons1}\\
    &&\delta^* = \frac{L}{2(L-1)\omega_3}\gamma_Y^2\gamma_X^{-2},\label{case3-cons2}\\
    &&\omega_1=\frac{\lambda_Y^{-1}-\gamma_Y\lambda_Y^{-2}}{2}-\omega_3\lambda_X^2\lambda_Y^{-2},\\
    &&\omega_3 = \frac{L}{2(LD-\lambda_X-(L-1)(\gamma_X-\gamma_X^2\gamma_Y^{-1}))}.\label{omega3-cons}
    \end{IEEEeqnarray}
    \end{subequations}
   Considering the fact that $\omega_1\geq 0$ yields the following constraint:
    \begin{IEEEeqnarray}{rCl}
    \omega_3\leq \frac{1}{2}\lambda_X^{-2}(\lambda_Y-\gamma_Y).\label{ineq2}
    \end{IEEEeqnarray}
Combining the above inequality with \eqref{omega3-cons}, we get
\begin{IEEEeqnarray}{rCl}
LD\geq L\lambda_X^2(\lambda_Y-\gamma_Y)^{-1}+(L-1)\gamma_X^2(\gamma_X^{-1}-\gamma_Y^{-1})+\lambda_X.\label{subcase2-condition2}
\end{IEEEeqnarray}
With a similar reason to the previous subcase (by considering distortion constraints \eqref{Dc-cons} and \eqref{Dc-cons2}), we also know that $D\in\mathcal{D}_1^c(\lambda_I)$. In summary, for this subcase, the distortion set is restricted to $\mathcal{D}_1^c(\lambda_I)$ while constraint~\eqref{subcase2-condition2} holds.   Plugging \eqref{case3-cons1} and \eqref{case3-cons2} into \eqref{obj-function} while considering \eqref{omega3-cons} gives the rate-distortion expression $\underline{\mathcal{R}}_2^c(D)$ defined in~\eqref{R1D}.

To sum up all of the above cases,  we have
\begin{IEEEeqnarray}{rCl}
\underline{\mathcal{R}}(D)=\left\{\begin{array}{ll}\overline{\mathcal{R}}(D),&\;\;\; D\in\mathcal{D}_1(\lambda_I),\\
\underline{\mathcal{R}}^c(D),&\;\;\;D\in\mathcal{D}_1^c(\lambda_I),\end{array}\right.\label{RDl-lg}
\end{IEEEeqnarray}
where $\underline{\mathcal{R}}^c(D)$ is defined in \eqref{Rc-def}.

Next, consider the case $\gamma_Y\geq \lambda_Y>0$, where we have $\lambda_W=\lambda_Y$. Thus, the objective function \eqref{general-obj-function} reduces to 
\begin{IEEEeqnarray}{rCl}
&&\hspace{1cm}f(\beta,\delta):=
\frac{L-1}{2}\log \frac{\gamma_Y^2}{(\gamma_Y-\lambda_Y)\beta+\lambda_Y\gamma_Y}+\frac{L}{2}\log \frac{\lambda_Y}{\delta},
\end{IEEEeqnarray}
subject to the following constraints:
\begin{subequations}\label{opt4}
\begin{IEEEeqnarray}{rCl}
&&0<\beta\leq \gamma_Y,\\
&&0<\delta \leq  (\beta^{-1}+\lambda_Y^{-1}-\gamma_Y^{-1})^{-1},\\
&&\lambda_X^2\lambda_Y^{-2}\delta+\lambda_X-\lambda_X^2\lambda_Y^{-1}+(L-1)(\gamma_X^2\gamma_Y^{-2}\beta+\gamma_X-\gamma_X^2\gamma_Y^{-1})\leq LD. 
\end{IEEEeqnarray}
 \end{subequations}
Then, the goal is to minimize $f(\beta,\delta)$ subject to the constraints \eqref{opt4}. The solution to this convex program can be obtained following similar steps to the case of $\lambda_Y\geq \gamma_Y$. Under the condition
\begin{IEEEeqnarray}{rCl}
 \lambda_X^2\lambda_Y^{-2}(\gamma_I^{-1}+\lambda_Y^{-1}-\gamma_Y^{-1})^{-1}-L\gamma_I\gamma_X^2\gamma_Y^{-2}&\leq& L\gamma_X^2\gamma_Y^{-1}(\gamma_Y\lambda_Y^{-1}-1)^{-1},\label{case4-cons}
\end{IEEEeqnarray}
the admissible distortion set is given by 
\begin{IEEEeqnarray}{rCl}
&&\mathcal{D}_2(\gamma_I):=\{D\in (d_{\min},\sigma_X^2)\colon  \nonumber\\&&\hspace{0.2cm} LD\leq L\gamma_X^2\gamma_Y^{-1}(\gamma_Y\lambda_Y^{-1}-1)^{-1}+(L-1)\gamma_X^2(\gamma_X^{-1}-\gamma_Y^{-1})+\lambda_X-\lambda_X^2\lambda_Y^{-1}+(2L-1)\gamma_{I}\gamma_X^2\gamma_Y^{-2},\nonumber\\
&&\hspace{0.2cm}LD=\lambda_X-\lambda_X^2\lambda_Y^{-1}+(L-1)(\gamma_X^2\gamma_Y^{-2}\gamma_I+\gamma_X-\gamma_X^2\gamma_Y^{-1})+\lambda_X^2\lambda_Y^{-2}(\gamma_{I}^{-1}+\lambda_Y^{-1}-\gamma_Y^{-1})^{-1}\},
\end{IEEEeqnarray}
where we have $\underline{\mathcal{R}}(D)=\overline{\mathcal{R}}(D)$. Moreover, under the condition
\begin{IEEEeqnarray}{rCl}
 \lambda_X^2\lambda_Y^{-2}(\gamma_I^{-1}+\lambda_Y^{-1}-\gamma_Y^{-1})^{-1}-L\gamma_I\gamma_X^2\gamma_Y^{-2}&>& L\gamma_X^2\gamma_Y^{-1}(\gamma_Y\lambda_Y^{-1}-1)^{-1},\label{case4-cons-complement}
\end{IEEEeqnarray}
the admissible distortion set is given by 
\begin{IEEEeqnarray}{rCl}
&&\mathcal{D}_2^c(\gamma_I):=\{D\in (d_{\min},\sigma_X^2)\colon  \nonumber\\&&\hspace{0.2cm} LD= L\gamma_X^2\gamma_Y^{-1}(\gamma_Y\lambda_Y^{-1}-1)^{-1}+(L-1)\gamma_X^2(\gamma_X^{-1}-\gamma_Y^{-1})+\lambda_X-\lambda_X^2\lambda_Y^{-1}+(2L-1)\gamma_{I}\gamma_X^2\gamma_Y^{-2},\nonumber\\
&&\hspace{0.2cm}LD<\lambda_X-\lambda_X^2\lambda_Y^{-1}+(L-1)(\gamma_X^2\gamma_Y^{-2}\gamma_I+\gamma_X-\gamma_X^2\gamma_Y^{-1})+\lambda_X^2\lambda_Y^{-2}(\gamma_{I}^{-1}+\lambda_Y^{-1}-\gamma_Y^{-1})^{-1}\},\footnote{The inequality constraint inside the set $\mathcal{D}_2^c(\gamma_I)$ is implied by the equality condition on $D$ and constraint~\eqref{case4-cons-complement}. A similar statement holds for the sets $\mathcal{D}_1(\lambda_I)$, $\mathcal{D}_2(\gamma_I)$ and $\mathcal{D}_1^c(\lambda_I)$. Both conditions on $D$ are included in the definition of these sets to show that $\mathcal{D}_1^c(\lambda_I)$ (resp. $\mathcal{D}_2^c(\gamma_I)$) is a complement of $\mathcal{D}_1(\lambda_I)$ (resp. $\mathcal{D}_2(\gamma_I)$).}\nonumber\\
\end{IEEEeqnarray}
where the lower bound takes the expression $\underline{\mathcal{R}}^c(D)$ defined in \eqref{Rchat-def}.
Thus, the case of $\gamma_Y\geq \lambda_Y$ can be summarized as follows:
\begin{IEEEeqnarray}{rCl}
\underline{\mathcal{R}}(D)=\left\{\begin{array}{ll}\overline{\mathcal{R}}(D),&\;\;\; D\in\mathcal{D}_2(\gamma_I),\\
\hat{\underline{\mathcal{R}}}^c(D),&\;\;\;D\in\mathcal{D}_2^c(\gamma_I).\end{array}\right.\label{RDl-gl}
\end{IEEEeqnarray}
After characterizing the lower bound under two complement sets for each of the cases $\lambda_Y\geq \gamma_Y$ and $\gamma_Y\geq \lambda_Y$, it just remains to explicitly determine the sets $\mathcal{D}_1(\lambda_I)$ and $\mathcal{D}_2(\gamma_I)$. With some straightforward calculations, it can be verified that
\begin{IEEEeqnarray}{rCl}
&&\mathcal{D}_1(\lambda_I)\nonumber\\&&\hspace{0.7cm}=\left\{ \begin{array}{ll} \{D\in (d_{\min},\sigma_X^2)\} &\hspace{0cm}\text{if}\;\; \lambda_X^2\gamma_X^{-2}\gamma_Y^2\lambda_Y^{-2}\geq\frac{L-1}{4L},\\  \{D\in (d_{\min},\sigma_X^2)\} &\hspace{0cm} \text{if}\;\; \lambda_X^2\gamma_X^{-2}\gamma_Y^2\lambda_Y^{-2}<\frac{L-1}{4L} \; \text{and}\; \mu_2\leq \frac{\gamma_Y}{\lambda_Y},\\ \{D\in (d_{\min}, \mathsf{D}_{\text{th},1})\}&\hspace{0cm} \text{if}\;\;\lambda_X^2\gamma_X^{-2}\gamma_Y^2\lambda_Y^{-2}<\frac{L-1}{4L}, \mu_1\leq \frac{\gamma_Y}{\lambda_Y} \;\text{and}\;\frac{\gamma_Y}{\lambda_Y}<\mu_2<1,\\ \{D\in (d_{\min},\mathsf{D}_{\text{th},1}) \cup
(\mathsf{D}_{\text{th},2},\sigma_X^2)\}&\text{if}\;\;\lambda_X^2\gamma_X^{-2}\gamma_Y^2\lambda_Y^{-2}<\frac{L-1}{4L},\;\mu_1>\frac{\gamma_Y}{\lambda_Y} \;\text{and}\;\mu_2<1 ,\\ \emptyset&\hspace{0cm}\text{if}\;\;\lambda_X^2\gamma_X^{-2}\gamma_Y^2\lambda_Y^{-2}<\frac{L-1}{4L},\;\mu_1=0 \;\text{and}\;\mu_2=1, \end{array}\right.\nonumber\\
\end{IEEEeqnarray}
and
\begin{IEEEeqnarray}{rCl}
&&\mathcal{D}_2(\gamma_I)\nonumber\\&&\hspace{0.7cm}=\left\{ \begin{array}{ll} \{D\in (d_{\min},\sigma_X^2)\} &\hspace{0cm}\text{if}\;\; \gamma_X^2\lambda_X^{-2}\lambda_Y^2\gamma_Y^{-2}\geq\frac{1}{4L},\\ 
\{D\in (d_{\min},\sigma_X^2)\} & \hspace{0cm}\text{if}\;\; \gamma_X^2\lambda_X^{-2}\lambda_Y^2\gamma_Y^{-2}<\frac{1}{4L} \;\text{and}\;\nu_2\leq \frac{\lambda_Y}{\gamma_Y},\\
\{D\in (d_{\min}, \hat{\mathsf{D}}_{\text{th},1})\}& \hspace{0cm}\text{if}\;\;\gamma_X^2\lambda_X^{-2}\lambda_Y^2\gamma_Y^{-2}<\frac{1}{4L}, \nu_1\leq \frac{\lambda_Y}{\gamma_Y} \;\text{and}\;\;\frac{\lambda_Y}{\gamma_Y}<\nu_2<1,\\ 
\{D\in (d_{\min},\hat{\mathsf{D}}_{\text{th},1})\cup (\hat{\mathsf{D}}_{\text{th},2},\sigma_X^2)\}&\hspace{0cm}\text{if}\;\;\gamma_X^2\lambda_X^{-2}\lambda_Y^2\gamma_Y^{-2}<\frac{1}{4L}, \nu_1>\frac{\lambda_Y}{\gamma_Y} \;\text{and}\;\nu_2<1,\\ \emptyset&\hspace{0cm}\text{if}\;\;\gamma_X^2\lambda_X^{-2}\lambda_Y^2\gamma_Y^{-2}<\frac{1}{4L}, \nu_1=0 \;\text{and}\;\nu_2=1.\end{array}\right.\nonumber\\
\end{IEEEeqnarray}
This completes the proof.

\section{Proof of Theorem \ref{upper-asym}}\label{upper-asym-proof}
First, notice that the distortion constraint in \eqref{distortion-ach} can be written as
\begin{IEEEeqnarray}{rCl}
(\lambda_X+(L-1)\gamma_X-LD)\lambda_Q^2+(\phi_1\gamma_Y+(L-1)\phi_2\lambda_Y-\phi_3(\gamma_Y+\lambda_Y))\lambda_Q-\phi_3\lambda_Y\gamma_Y=0,\nonumber\\\label{asym-equation}
\end{IEEEeqnarray}
where $\phi_1 := \lambda_X^2\lambda_Y^{-1}$, $\phi_2 := \gamma_X^2\gamma_Y^{-1}$ and $\phi_3 := LD+\phi_1+(L-1)\phi_2-(\lambda_X+(L-1)\gamma_X)$. 
The equation in \eqref{asym-equation} can be equivalently written as 
\begin{IEEEeqnarray}{rCl}
a\lambda_Q^2+b\lambda_Q+c=0,
\end{IEEEeqnarray}
where $a:=(\sigma_X^2-D)L$, $b:= g_1L^2+g_2L$ and $c:= h_1 L^2+h_2L$ and 
\begin{IEEEeqnarray}{rCl}
g_1 &:=& \rho_X\rho_Z\sigma_X^2\sigma_Z^2+(\rho_X\sigma_X^2+\rho_Z\sigma_Z^2)(\gamma_X-D),\label{g1-def}\\
g_2 &:=& \sigma_X^2(\gamma_Z+\gamma_Y)-\rho_X\sigma_X^2\gamma_X-2\gamma_YD,\\
h_1 &:=& \rho_X\rho_Z\sigma_X^2\sigma_Z^2\gamma_Y+(\rho_X\sigma_X^2+\rho_Z\sigma_Z^2)(\gamma_X\gamma_Z-\gamma_YD)=\gamma_Y(\rho_X\sigma_X^2+\rho_Z\sigma_Z^2)(d_{\min}^{\infty}-D),\nonumber\\\label{h1-def}\\
h_2 &:=& \rho_X\sigma_X^2\gamma_Z^2+\rho_Z\sigma_Z^2\gamma_X^2+\gamma_X\gamma_Z\gamma_Y-\gamma_Y^2D.
\end{IEEEeqnarray}
We consider three different cases based on the value of $g_1$.

\textbf{Case1 ($g_1>0$):} In this case, we have
\begin{IEEEeqnarray}{rCl}
\lambda_Q &=& \frac{-b+\sqrt{b^2-4ac}}{2a}\\
&=& \frac{-b+b\sqrt{1-\frac{4ac}{b^2}}}{2a}\\
&=& \frac{-b+b(1-\frac{2ac}{b^2}-\frac{2a^2c^2}{b^4}+O(\frac{1}{L^3}))}{2a}\label{just1}\\
&=& -\frac{c}{b}-\frac{ac^2}{b^3}+O\left(\frac{1}{L^2}\right)\\
&=& -\frac{h_1L+h_2}{g_1L+g_2}-\frac{(\sigma_X^2-D)(h_1L+h_2)^2}{(g_1L+g_2)^3}+O\left(\frac{1}{L^2}\right)\\
&=&-\frac{h_1L+h_2}{g_1L}\left(1-\frac{g_2}{g_1L}+O(\frac{1}{L^2})\right)-\frac{(\sigma_X^2-D)h_1^2}{g_1^3L}+O\left(\frac{1}{L^2}\right)\label{just2}\\
&=& -\frac{h_1}{g_1}-\left(\frac{h_2}{g_1}-\frac{g_2h_1}{g_1^2}+\frac{(\sigma_X^2-D)h_1^2}{g_1^3}\right)\frac{1}{L}+O\left(\frac{1}{L^2}\right)\\
&:=&\eta_1+\frac{\eta_2}{L}+O\left(\frac{1}{L^2}\right),
\end{IEEEeqnarray}
where \eqref{just1} follows because $\sqrt{1+x}= 1+\frac{1}{2}x-\frac{1}{8}x^2+O(x^3)$ and \eqref{just2} follows because $\frac{1}{1+x}= 1-x+O(x^2)$. Now, plugging the above into \eqref{ach-rate} yields 
\begin{IEEEeqnarray}{rCl}
&&\hspace{-1cm} \frac{1}{2}\log\frac{\lambda_Y+\lambda_Q}{\lambda_Q}+\frac{L-1}{2}\log \frac{\gamma_Y+\lambda_Q}{\lambda_Q}\label{rate-expression}\\
&=&\frac{1}{2}\log \frac{\lambda_Y+\eta_1+\frac{\eta_2}{L}+O(\frac{1}{L^2})}{\eta_1+\frac{\eta_2}{L}+O(\frac{1}{L^2})}\nonumber\\&&\hspace{1cm}+\frac{L-1}{2}\log \frac{\gamma_Y+\eta_1+\frac{\eta_2}{L}+O(\frac{1}{L^2})}{\eta_1+\frac{\eta_2}{L}+O(\frac{1}{L^2})}\\
&=&\frac{1}{2}\log \frac{(1+(L-1)\rho_Y)\sigma_Y^2+\eta_1+\frac{\eta_2}{L}+O(\frac{1}{L^2})}{\eta_1+\frac{\eta_2}{L}+O(\frac{1}{L^2})}\nonumber\\&&\hspace{1cm}+\frac{L-1}{2}\log \frac{\gamma_Y+\eta_1+\frac{\eta_2}{L}+O(\frac{1}{L^2})}{\eta_1+\frac{\eta_2}{L}+O(\frac{1}{L^2})}\\
&=&\frac{1}{2}\log \left(\frac{L\rho_Y\sigma_Y^2}{\eta_1}+O(1)\right)\nonumber\\
&&\hspace{0.1cm}+\frac{L-1}{2}\log\left(\left(\frac{\gamma_Y+\eta_1}{\eta_1}+\frac{\eta_2}{L\eta_1}+O\left(\frac{1}{L^2}\right)\right)\left(1-\frac{\eta_2}{L\eta_1}+O\left(\frac{1}{L^2}\right)\right)\right)\label{just3}\\
&=&\frac{1}{2}\log\left(\frac{L\rho_Y\sigma_Y^2}{\eta_1}+O(1)\right)+\frac{L-1}{2}\log\left(\frac{\gamma_Y+\eta_1}{\eta_1}-\frac{\eta_2\gamma_Y}{L\eta_1^2}+O\left(\frac{1}{L^2}\right)\right)\\
&=&\frac{1}{2}\log L+\frac{1}{2}\log \frac{\rho_Y\sigma_Y^2}{\eta_1+\gamma_Y}+\frac{L}{2}\log \frac{\eta_1+\gamma_Y}{\eta_1}-\frac{\eta_2\gamma_Y}{2\eta_1(\eta_1+\gamma_Y)}+O\left(\frac{1}{L}\right)\label{just4}\\
&=&\frac{1}{2}\log L+\frac{1}{2}\log \frac{\rho_X\sigma_X^2+\rho_Z\sigma_Z^2}{\eta_1+\gamma_Y}+\frac{L}{2}\log \frac{\eta_1+\gamma_Y}{\eta_1}-\frac{\eta_2\gamma_Y}{2\eta_1(\eta_1+\gamma_Y)}+O\left(\frac{1}{L}\right),\label{app-final2}
\end{IEEEeqnarray}
where \eqref{just3} follows because $\frac{1}{1+x}= 1-x+O(x^2)$ and \eqref{just4} follows because $\log(1+x)=x+O(x^2)$.
With some straightforward calculations, we can show that each term of the above expression can be written as follows:
\begin{subequations}\label{app-final2-each}
\begin{IEEEeqnarray}{rCl}
&&\frac{1}{2}\log \frac{\rho_X\sigma_X^2+\rho_Z\sigma_Z^2}{\eta_1+\gamma_Y}=\frac{1}{2}\log \frac{\rho_X\rho_Z\sigma_X^2\sigma_Z^2+(\rho_X\sigma_X^2+\rho_Z\sigma_Z^2)(\gamma_X-D)}{\gamma_X^2},\\
&&\frac{L}{2}\log \frac{\eta_1+\gamma_Y}{\eta_1}=  \frac{L}{2}\log \frac{(\rho_X\sigma_X^2+\rho_Z\sigma_Z^2)\gamma_X^2}{(\rho_X\sigma_X^2+\rho_Z\sigma_Z^2)(\gamma_YD-\gamma_X\gamma_Z)-\rho_X\rho_Z\gamma_Y\sigma_X^2\sigma_Z^2},\\
&&-\frac{\gamma_Y\eta_2}{2\eta_1(\eta_1+\gamma_Y)} =  \frac{\gamma_Y (\sigma_X^2\rho_Z\sigma_Z^2-(\rho_X\sigma_X^2+\rho_Z\sigma_Z^2)D)^2}{2(\rho_X\rho_Z\sigma_X^2\sigma_Z^2+(\rho_X\sigma_X^2+\rho_Z\sigma_Z^2)(\gamma_X-D))(\rho_X\sigma_X^2+\rho_Z\sigma_Z^2)\gamma_Y(D-d_{\min}^{\infty})}\nonumber\\
&&=\frac{\gamma_Y (\sigma_X^2\rho_Z\sigma_Z^2-(\rho_X\sigma_X^2+\rho_Z\sigma_Z^2)D)^2}{2(\rho_X\rho_Z\sigma_X^2\sigma_Z^2+(\rho_X\sigma_X^2+\rho_Z\sigma_Z^2)(\gamma_X-D)) ( (\rho_X\sigma_X^2+\rho_Z\sigma_Z^2)(\gamma_Y D-\gamma_X \gamma_Z)-\rho_X\rho_Z\gamma_Y\sigma_X^2\sigma_Z^2) }.\nonumber\\
\end{IEEEeqnarray}
\end{subequations}
Moreover, notice that $g_1>0$ and $D>d_{\min}^{\infty}$ implies $\rho_X\sigma_X^2+\rho_Z\sigma_Z^2>0$ and $\eta_1>0$ from \eqref{g1-def} and \eqref{h1-def}. Considering these conditions, \eqref{app-final2}--\eqref{app-final2-each} and simplifying the terms, we get the first clause of \eqref{approx-upp2}.

\textbf{Case 2 ($g_1=0$):} We consider two different subcases.

\underline{\textit{Subcase 1 ($\rho_X\sigma_X^2+\rho_Z\sigma_Z^2=0$):}}  The distortion constraint in~\eqref{distortion-ach} simplifies to 
\begin{IEEEeqnarray}{rCl}
L\sigma_X^2-\frac{L\sigma_X^4}{\sigma_X^2+\sigma_Z^2+\lambda_Q}=LD, 
\end{IEEEeqnarray}
or equivalently,
\begin{IEEEeqnarray}{rCl}
\lambda_Q=\frac{\sigma_X^4}{\sigma_X^2-D}-\sigma_X^2-\sigma_Z^2.
\end{IEEEeqnarray}
Plugging the above solution in~\eqref{ach-rate}, we get the rate-distortion expression in \eqref{approx-upp1}.

\underline{\textit{Subcase 2 ($\rho_X\sigma_X^2+\rho_Z\sigma_Z^2>0$):}} In this case, we have

\begin{IEEEeqnarray}{rCl}
\lambda_Q &=& \frac{-g_2L+g_2L\sqrt{1-4\frac{(\sigma_X^2-D)(h_1L+h_2)}{g_2^2}}}{2L(\sigma_X^2-D)}\\
&=& \frac{-g_2L+L^{\frac{3}{2}}\sqrt{-4(\sigma_X^2-D)h_1}\sqrt{1+\frac{g_2^2-4(\sigma_X^2-D)h_2}{-4(\sigma_X^2-D)h_1L}}}{2L(\sigma_X^2-D)}\\
&=& \frac{-g_2L+L^{\frac{3}{2}}\sqrt{-4(\sigma_X^2-D)h_1}(1-\frac{g_2^2-4(\sigma_X^2-D)h_2}{8(\sigma_X^2-D)h_1L}+O(\frac{1}{L^2}))}{2L(\sigma_X^2-D)}\\
&=&\sqrt{-\frac{h_1L}{\sigma_X^2-D}}-\frac{g_2}{2(\sigma_X^2-D)}+\frac{g_2^2-4(\sigma_X^2-D)h_2}{8\sqrt{-(\sigma_X^2-D)^3h_1L}}+O(\frac{1}{L^{\frac{3}{2}}})\\
&:=& \alpha_1\sqrt{L}+\alpha_2+O(\frac{1}{\sqrt{L}}).\label{just9}
\end{IEEEeqnarray}
Moreover, the condition $\rho_X\sigma_X^2+\rho_Z\sigma_Z^2>0$ together with $\sigma_X^2>D>d_{\min}^{\infty}$ and $g_1=0$ implies $\rho_X>0$, $\gamma_X>0$ and $\alpha_1>0$. Then, we get the following:
\begin{IEEEeqnarray}{rCl}
&&D=\frac{\rho_X\rho_Z\sigma_X^2\sigma_Z^2}{\rho_X\sigma_X^2+\rho_Z\sigma_Z^2}+\gamma_Y,\\
&&\sigma_X^2-D=\frac{\rho_X^2\sigma_X^4}{\rho_X\sigma_X^2+\rho_Z\sigma_Z^2},\\
&&h_1 =-(\rho_X\sigma_X^2+\rho_Z\sigma_Z^2)\gamma_X^2,\\
&&g_2 = \frac{-\rho_Z\gamma_X\sigma_X^2\sigma_Z^2-\rho_X\gamma_X\gamma_Y\sigma_X^2+\rho_X^2\sigma_X^4\gamma_Z+\rho_X\gamma_Z\sigma_X^4-\rho_X\rho_Z\gamma_X\sigma_X^2\sigma_Z^2}{\rho_X\sigma_X^2+\rho_Z\sigma_Z^2},\\
&&\alpha_1=\frac{(\rho_X\sigma_X^2+\rho_X\sigma_Z^2)\gamma_X}{\rho_X\sigma_X^2},\label{alpha-def1000}\\
&&\alpha_2 = \frac{\rho_Z\gamma_X\sigma_X^2\sigma_Z^2+\rho_X\gamma_X\gamma_Y\sigma_X^2-\rho_X^2\sigma_X^4\gamma_Z-\rho_X\gamma_Z\sigma_X^4+\rho_X\rho_Z\gamma_X\sigma_X^2\sigma_Z^2}{2\rho_X^2\sigma_X^4}.
\end{IEEEeqnarray}
Now, we simplify each term of the rate in \eqref{ach-rate}. Consider the first term of \eqref{ach-rate} as follows:
\begin{IEEEeqnarray}{rCl}
\frac{1}{2}\log \frac{\lambda_Y+\lambda_Q}{\lambda_Q} &=& \frac{1}{2}\log \frac{\lambda_Y}{\lambda_Q}+\frac{1}{2}\log \frac{\lambda_Y+\lambda_Q}{\lambda_Y}\\
&=& \frac{1}{2}\log \frac{L(\rho_X\sigma_X^2+\rho_Z\sigma_Z^2)+\gamma_Y}{\alpha_1\sqrt{L}+O\left(1\right)}+\frac{1}{2}\log \frac{\lambda_Y+\lambda_Q}{\lambda_Y}\label{just7}\\
&=& \frac{1}{2}\log \frac{L(\rho_X\sigma_X^2+\rho_Z\sigma_Z^2)+\gamma_Y}{\alpha_1\sqrt{L}+O\left(1\right)}+O\left(\frac{1}{\sqrt{L}}\right)\label{just6}\\
&=& \frac{1}{4}\log L+\frac{1}{2}\log \frac{\rho_X\sigma_X^2+\rho_Z\sigma_Z^2}{\alpha_1}+O\left(\frac{1}{\sqrt{L}}\right)\\
&=& \frac{1}{4}\log L+\frac{1}{2}\log \rho_X\gamma_X^{-1}\sigma_X^2+O\left(\frac{1}{\sqrt{L}}\right),\label{just8}
\end{IEEEeqnarray}
where \eqref{just7} follows from the definition of $\lambda_Y$ in \eqref{lamy-def} and the definition of $\lambda_Q$ in \eqref{just9}, \eqref{just6} follows because $\frac{\lambda_Q}{\lambda_Y}=O(\frac{1}{\sqrt{L}})$ and $\log(1+x)=O(x)$, \eqref{just8} follows from the definition of $\alpha_1$ in \eqref{alpha-def1000}.

The second term of \eqref{ach-rate} can be simplified as follows:
\begin{IEEEeqnarray}{rCl}
\frac{L-1}{2}\log \frac{\gamma_Y+\lambda_Q}{\lambda_Q} &=& \frac{L-1}{2}\log (1+\frac{\gamma_Y}{\lambda_Q})\\
&=& \frac{L-1}{2} \left(\frac{\gamma_Y}{\lambda_Q}-\frac{\gamma_Y^2}{2\lambda_Q^2}+O\left(\frac{1}{L^{\frac{3}{2}}}\right)\right)\label{just10}\\
&=& \frac{L-1}{2} \left(\frac{\gamma_Y}{\alpha_1\sqrt{L}+\alpha_2+O(\frac{1}{\sqrt{L}})}-\frac{\gamma_Y^2}{2(\alpha_1\sqrt{L}+O(1))^2}+O\left(\frac{1}{L^{\frac{3}{2}}}\right)\right)\nonumber\\
&=&\frac{L-1}{2} \Big(\frac{\gamma_Y}{\alpha_1\sqrt{L}}\left(1-\frac{\alpha_2}{\alpha_1\sqrt{L}}+O\left(\frac{1}{L}\right)\right)-\frac{\gamma_Y^2}{2\alpha_1^2L}\left(1+O\left(\frac{1}{\sqrt{L}}\right)\right)\nonumber\\&&\hspace{1.5cm}+O\left(\frac{1}{L^{\frac{3}{2}}}\right)\Big)\label{just11}\\
&=& \frac{\gamma_Y\sqrt{L}}{2\alpha_1}-\frac{\gamma_Y(\gamma_Y+2\alpha_2)}{4\alpha_1^2}+O\left(\frac{1}{\sqrt{L}}\right)\\
&=&\frac{\rho_X\gamma_Y\sigma_X^2\sqrt{L}}{2\gamma_X(\rho_X\sigma_X^2+\rho_Z\sigma_Z^2)}-\frac{\gamma_Y(\rho_X\sigma_X^4(\gamma_X-\rho_X\gamma_Z)+(1+\rho_X)\rho_Z\sigma_X^2\sigma_Z^2\gamma_X)}{4(\rho_X\sigma_X^2+\rho_Z\sigma_Z^2)^2\gamma_X^2}\nonumber\\&&\hspace{0.5cm}+O\left(\frac{1}{\sqrt{L}}\right),\label{just200}
\end{IEEEeqnarray}
where \eqref{just10} follows because $\frac{\gamma_Y}{\lambda_Q}=O\left(\frac{1}{\sqrt{L}}\right)$ and $\log(1+x)=x-\frac{1}{2}x^2+O(x^3)$, \eqref{just11} follows because $\frac{1}{1+x}=1-x+O(x^2)$. Considering the fact that $g_1=0$, using approximations \eqref{just8} and \eqref{just200} and simplifying the terms, we get the second clause of \eqref{approx-upp2}.

\textbf{Case 3 ($g_1<0$):} Here, we have
\begin{IEEEeqnarray}{rCl}
\lambda_Q &=& \frac{-g_1L^2-g_2L+\sqrt{(g_1L^2+g_2L)^2-4L(\sigma_X^2-D)(h_1L^2+h_2L)}}{2L(\sigma_X^2-D)}\\
&=& \frac{-g_1L^2-g_2L+\sqrt{g_1^2L^4+(2g_1g_2-4(\sigma_X^2-D)h_1)L^3+(g_2^2-4(\sigma_X^2-D)h_2)L^2}}{2L(\sigma_X^2-D)}\\
&=& \frac{-g_1L^2-g_2L-g_1L^2(1+(g_1g_2-2(\sigma_X^2-D)h_1)\frac{1}{g_1^2L}+O(\frac{1}{L^2}))}{2L(\sigma_X^2-D)}\label{just12}\\
&=&-\frac{g_1}{\sigma_X^2-D}L-\frac{g_2-(\sigma_X^2-D)h_1}{\sigma_X^2-D}+O\left(\frac{1}{L}\right)\\
&=& \frac{(\gamma_X-D)(\rho_X\sigma_X^2+\rho_Z\sigma_Z^2)-\rho_X\rho_Z\sigma_X^2\sigma_Z^2}{\sigma_X^2-D}+O(1)\\
&:=& \beta_1L+O(1),
\end{IEEEeqnarray}
where \eqref{just12} follows because $\sqrt{1+x}=1+\frac{1}{2}x+O(x^2)$. We then use the above approximation to calculate each term of the rate in \eqref{ach-rate} as follows:
\begin{IEEEeqnarray}{rCl}
\frac{1}{2}\log \frac{\lambda_Y+\lambda_Q}{\lambda_Q} &=& \frac{1}{2}\log \frac{\rho_X\sigma_X^2+\rho_Z\sigma_Z^2+\beta_1}{\beta_1}+O\left(\frac{1}{L}\right)\\
&=& \frac{1}{2}\log \frac{\rho_X^2\sigma_X^4}{(\rho_X\sigma_X^2+\rho_Z\sigma_Z^2)(D-\gamma_X)-\rho_X\rho_Z\sigma_X^2\sigma_Z^2}+O\left(\frac{1}{L}\right),\label{just201}
\end{IEEEeqnarray}
and 
\begin{IEEEeqnarray}{rCl}
\frac{L-1}{2}\log \frac{\gamma_Y+\lambda_Q}{\lambda_Q} &=& \frac{L-1}{2}\left(\frac{\gamma_Y}{\lambda_Q}+O(\frac{1}{L^2})\right)\\
&=& \frac{\gamma_Y}{2\beta_1}+O\left(\frac{1}{L}\right)\\
&=&\frac{\gamma_Y(\sigma_X^2-D)}{2(\rho_X\sigma_X^2+\rho_Z\sigma_Z^2)(D-\gamma_X)-2\rho_X\rho_Z\sigma_X^2\sigma_Z^2}+O\left(\frac{1}{L}\right).\label{just202}
\end{IEEEeqnarray}
Considering the fact that $g_1<0$, using approximations \eqref{just201} and \eqref{just202} and simplifying the terms, we get the third clause of \eqref{approx-upp2}. This concludes the proof.
\section{Proof of Theorem \ref{lower-asym-thm}}\label{lower-asym-thm-proof}

First, notice that $\rho_X,\rho_Z\in [0,1]$ implies $\lambda_{Y}\geq \gamma_{Y}$.  We consider four different cases.

\textbf{Case 1 ($\rho_X\sigma_X^2+\rho_Z\sigma_Z^2=0$)}: In this case, the condition $\lambda_X^2\gamma_Y^2\geq \frac{L-1}{4L}\gamma_X^2\lambda_Y^2$ is satisfied trivially for all $L$. So, we are under the first condition of Theorem~\ref{conv-new}, and consequently 
\begin{IEEEeqnarray}{rCl}
\underline{\mathcal{R}}(D)=\overline{\mathcal{R}}(D)=\overline{\mathcal{R}}^{\infty}(D).
\end{IEEEeqnarray}
This yields the first condition of Theorem~\eqref{lower-asym-thm}, where the rate-distortion expression is given by~\eqref{lower-thm-cond1}. 

\textbf{Case 2 ($\rho_X\sigma_X^2+\rho_Z\sigma_Z^2>0$, $\rho_X>0$, $\xi\geq \frac{1}{2}$)}: In this case, we are under the first condition of Theorem~\ref{conv-new}. This can be readily verified when $\gamma_X=0$. When $\gamma_X>0$, we have
\begin{IEEEeqnarray}{rCl}
\lambda_X^2\lambda_Y^{-2}\gamma_X^{-2}\gamma_Y^2&=&\frac{(1+(L-1)\rho_X)^2(1-\rho_Y)^2}{(1+(L-1)\rho_Y)^2(1-\rho_X)^2}\\
&=&\xi^2+\frac{2\xi^2(\rho_Y-\rho_X)}{\rho_X\rho_YL}+O\left(\frac{1}{L^2}\right)\\
&\geq &\frac{1}{4}\quad\mbox{for all sufficiently large }L\label{xi-argument}\\
&\geq &\frac{L-1}{4L},
\end{IEEEeqnarray}
where \eqref{xi-argument} can be verified by considering $\xi=\frac{1}{2}$ (which implies $\rho_Y>\rho_X$) and  $\xi>\frac{1}{2}$ separately.
In summary, the analysis of this case yields~\eqref{rate-final-lower-2nd-condition}.

\textbf{Case 3 ($\rho_X\sigma_X^2+\rho_Z\sigma_Z^2>0$, $\rho_X>0$, $\xi< \frac{1}{2}$)}: In this case, we are under the third condition of Theorem~\ref{conv-new}. This is because of the fact that $\mu_2<1$, 
\begin{IEEEeqnarray}{rCl}
\lambda_X^2\lambda_Y^{-2}\gamma_X^{-2}\gamma_Y^2
&=&\xi^2+O\left(\frac{1}{L}\right)\\
&<&\frac{L-1}{4L}\quad\mbox{for all sufficiently large }L,
\end{IEEEeqnarray}
and 
\begin{IEEEeqnarray}{rCl}
\mu_1 &=& \frac{1}{2}-\frac{1}{2}\sqrt{1-\frac{4L}{L-1}\lambda_X^2\lambda_Y^{-2}\gamma_X^{-2}\gamma_Y^2}\\
&=&\frac{1}{2}-\frac{1}{2}\sqrt{1-4\xi^2}+O\left(\frac{1}{L}\right)\\
&> &\frac{\gamma_Y}{\lambda_Y}\quad\mbox{for all sufficiently large }L,
\end{IEEEeqnarray}
where the last inequality follows because $\frac{\gamma_Y}{\lambda_Y}=O\left(\frac{1}{L}\right)$. Thus, we continue with approximating $\mathsf{D}_{\text{th},1}$, $\mathsf{D}_{\text{th},2}$ and the rate-distortion expressions. 
We approximate $\mathsf{D}_{\text{th},1}$ and $\mathsf{D}_{\text{th},2}$ for large $L$ as follows: 
\begin{IEEEeqnarray}{rCl}
\mathsf{D}_{\text{th},1}&=&\frac{\rho_X\rho_Z\sigma_X^2\sigma_Z^2}{\rho_X\sigma_X^2+\rho_Z\sigma_Z^2}+\gamma_X-\frac{1+\sqrt{1-4\xi^2}}{2}\gamma_X^2\gamma_Y^{-1}+O\left(\frac{1}{L}\right)\\
&=&\mathsf{D}^{\infty}_{\text{th},1}+O\left(\frac{1}{L}\right),
\end{IEEEeqnarray}
and
\begin{IEEEeqnarray}{rCl}
\mathsf{D}_{\text{th},2}&=&\frac{\rho_X\rho_Z\sigma_X^2\sigma_Z^2}{\rho_X\sigma_X^2+\rho_Z\sigma_Z^2}+\gamma_X-\frac{1-\sqrt{1-4\xi^2}}{2}\gamma_X^2\gamma_Y^{-1}+O\left(\frac{1}{L}\right)\\
&=&\mathsf{D}^{\infty}_{\text{th},2}+O\left(\frac{1}{L}\right).
\end{IEEEeqnarray}
Now, it remains to approximate the rate-distortion expressions. In the intervals $D<\mathsf{D}^{\infty}_{\text{th},1}$ and $D>\mathsf{D}^{\infty}_{\text{th},2}$, $\overline{\mathcal{R}}(D)$ can be approximated as in Theorem~\ref{upper-asym}, which leads to the expression in~\eqref{approx-upp2}. In the interval $\mathsf{D}^{\infty}_{\text{th},1}<D<\mathsf{D}^{\infty}_{\text{th},2}$, we need to approximate $\underline{\mathcal{R}}^c(D)$. For the rate-distortion expression $\underline{\mathcal{R}}^c(D)$, notice that the second clause of \eqref{Rc-def} is not active for large $L$ since 
\begin{IEEEeqnarray}{rCl}
L\lambda_X^2(\lambda_Y-\gamma_Y)^{-1}+(L-1)\gamma_X^2(\gamma_X^{-1}-\gamma_Y^{-1})+\lambda_X = L^2\rho_X^2\rho_Y^{-1}\sigma_X^4\sigma_Y^{-2}+O(L)>LD.\label{Rc-condition}
\end{IEEEeqnarray}
Thus, we need to approximate $\underline{\mathcal{R}}_1^c(D)$ defined in \eqref{R2Dl} for large $L$. 
Consider the following term in the first logarithm. We have
\begin{IEEEeqnarray}{rCl}
&&LD-\lambda_X-(L-1)(\gamma_X-\gamma_X^2\gamma_Y^{-1})+\lambda_X^2\lambda_Y^{-2}(\lambda_Y+(\gamma_Y^{-1}-\lambda_Y^{-1})^{-1})\nonumber\\
&&\hspace{0.5cm}=L(D-\rho_X\sigma_X^2-(\gamma_X-\gamma_X^2\gamma_Y^{-1})+\rho_X^2\rho_Y^{-1}\sigma_X^4\sigma_Y^{-2})+\nonumber\\&&\hspace{1cm}(2\rho_X\rho_Y^{-1}(1-\rho_X)\sigma_X^4\sigma_Y^{-2}-\gamma_X^2\gamma_Y^{-1})+O\left(\frac{1}{L}\right)\nonumber\\
&&\hspace{0.5cm}:=LA+B+O\left(\frac{1}{L}\right).
\end{IEEEeqnarray}
Thus, plugging the above into $\underline{\mathcal{R}}_1^c(D)$  in \eqref{R2Dl}, we can approximate the first logarithm as follows:
\begin{IEEEeqnarray}{rCl}
&&\hspace{-1.5cm}\frac{L+1}{2}\log \frac{(L+1)\gamma_Y^{-1}\gamma_X^2}{LA+B+O\left(\frac{1}{L}\right)}\nonumber\\&=&\frac{L+1}{2}\log \frac{\gamma_Y^{-1}\gamma_X^2}{A+\frac{1}{L+1}(-A+B)}\\
&=&\frac{L+1}{2}\log \frac{\gamma_Y^{-1}\gamma_X^2}{A}+\frac{A-B}{2A}+O\left(\frac{1}{L}\right)\\
&=&\frac{L+1}{2}\log\frac{\gamma_Y^{-1}\gamma_X^2}{D-\rho_X\sigma_X^2-(\gamma_X-\gamma_X^2\gamma_Y^{-1})+\rho_X^2\rho_Y^{-1}\sigma_X^4\sigma_Y^{-2}}\nonumber\\
&&\hspace{0.5cm}+\frac{1}{2}\frac{D+2\gamma_X^2\gamma_Y^{-1}-\sigma_X^2+\rho_X(3\rho_X-2)\rho_Y^{-1}\sigma_X^4\sigma_Y^{-2}}{D-\rho_X\sigma_X^2-(\gamma_X-\gamma_X^2\gamma_Y^{-1})+\rho_X^2\rho_Y^{-1}\sigma_X^4\sigma_Y^{-2}}+O\left(\frac{1}{L}\right)\\
&=&\frac{L+1}{2}\log \frac{\rho_X\sigma_X^2+\rho_Z\sigma_Z^2\gamma_Y^{-1}\gamma_X^2}{(\rho_X\sigma_X^2+\rho_Z\sigma_Z^2)(D-(\gamma_X-\gamma_X^2\gamma_Y^{-1}))-\rho_X\rho_Z\sigma_X^2\sigma_Z^2}\nonumber\\
&&\hspace{0.5cm}+\frac{1}{2}\frac{(\rho_X\sigma_X^2+\rho_Z\sigma_Z^2)(D+2(1-\xi)\gamma_X^2\gamma_Y^{-1}-\gamma_X)-\rho_X\rho_Z\sigma_X^2\sigma_Z^2}{(\rho_X\sigma_X^2+\rho_Z\sigma_Z^2)(D-(\gamma_X-\gamma_X^2\gamma_Y^{-1}))-\rho_X\rho_Z\sigma_X^2\sigma_Z^2}+O\left(\frac{1}{L}\right).\label{RDl2}
\end{IEEEeqnarray}
The second logarithm of \eqref{R2Dl} can  be approximated as follows:
\begin{IEEEeqnarray}{rCl}
\frac{1}{2}\log \lambda_X^2\gamma_X^{-2}(\lambda_Y\gamma_Y^{-1}-1)^{-1}=\frac{1}{2}\log L+\frac{1}{2}\log \left( \left(\frac{\rho_X}{1-\rho_X}\right)^2\left(\frac{1-\rho_Y}{\rho_Y}\right)\right)+O\left(\frac{1}{L}\right).\label{RDl2b}
\end{IEEEeqnarray}
The third logarithm of \eqref{R2Dl} can also be approximated as follows:
\begin{IEEEeqnarray}{rCl}
\frac{L}{2}\log \left(1-\frac{1}{L}\right)=-\frac{1}{2}+O\left(\frac{1}{L^2}\right).
\end{IEEEeqnarray}
Plugging \eqref{RDl2} and \eqref{RDl2b} into \eqref{R2Dl} yields  
\begin{IEEEeqnarray}{rCl}
\underline{\mathcal{R}}^c_1(D)&=&\frac{L+1}{2}\log \frac{\gamma_Y^{-1}\gamma_X^2}{(\rho_X\sigma_X^2+\rho_Z\sigma_Z^2)(D-(\gamma_X-\gamma_X^2\gamma_Y^{-1}))-\rho_X\rho_Z\sigma_X^2\sigma_Z^2}+\frac{1}{2}\log L\nonumber\\&&\hspace{0.5cm}+\frac{1}{2}\frac{(1-2\xi)\gamma_X^2\gamma_Y^{-1}}{(\rho_X\sigma_X^2+\rho_Z\sigma_Z^2)(D-(\gamma_X-\gamma_X^2\gamma_Y^{-1}))-\rho_X\rho_Z\sigma_X^2\sigma_Z^2}\nonumber\\&&\hspace{0.5cm}+\frac{1}{2}\log \left( \left(\frac{\rho_X}{1-\rho_X}\right)^2\left(\frac{1-\rho_Y}{\rho_Y}\right)\right)+O\left(\frac{1}{L}\right)\nonumber\\
&&=\underline{\mathcal{R}}_1^{\infty}(D).
\end{IEEEeqnarray}
The above expression can be further simplified to~\eqref{R1-lower-new}. Moreover, the two boundary points $D=\mathsf{D}^{\infty}_{\text{th},1}$ and $D=\mathsf{D}^{\infty}_{\text{th},2}$ can be easily handled by considering the fact that $\overline{\mathcal{R}}_1^{\infty}(\mathsf{D}^{\infty}_{\text{th},1})=\underline{\mathcal{R}}_1^{\infty}(\mathsf{D}^{\infty}_{\text{th},1})$ and $\overline{\mathcal{R}}_1^{\infty}(\mathsf{D}^{\infty}_{\text{th},2})=\underline{\mathcal{R}}_1^{\infty}(\mathsf{D}^{\infty}_{\text{th},2})$. In summary, the analysis of this case yields~\eqref{rate-final-lower-3rd-condition}.

\textbf{Case 4 ($\rho_X\sigma_X^2+\rho_Z\sigma_Z^2>0$ and $\rho_X=0$)}: In this case, we are under the second condition of Theorem~\ref{conv-new} since 
\begin{IEEEeqnarray}{rCl}
\mu_1 &=&\frac{1}{2}-\frac{1}{2}\sqrt{1-\frac{4L}{L-1}\lambda_Y^{-2}\gamma_Y^2}\\
&=&\frac{1}{2}-\frac{1}{2}\sqrt{1-\frac{4L}{L-1}\left(\frac{(1-\rho_Y)^2}{L^2\rho_Y^2}+O\left(\frac{1}{L^3}\right)\right)}\\
&=& \frac{(1-\rho_Y)^2}{L^2\rho_Y^2}+O\left(\frac{1}{L^3}\right)\\
&\leq &\frac{\gamma_Y}{\lambda_Y}\quad\text{for all sufficiently large}\;L,\label{eq:dueto1}
\end{IEEEeqnarray}
and
\begin{IEEEeqnarray}{rCl}
1&>&\mu_2 \\
&=& 1-\frac{(1-\rho_Y)^2}{L^2\rho_Y^2}+O\left(\frac{1}{L^3}\right)\\
&>& \frac{\gamma_Y}{\lambda_Y}\quad\text{for all sufficiently large}\;L,\label{eq:dueto2}
\end{IEEEeqnarray}
and
\begin{IEEEeqnarray}{rCl}
\lambda_X^2\lambda^{-2}_Y\gamma^{-2}_X\gamma_Y^2&=&O(\frac{1}{L^2})\\
&<&\frac{L-1}{4L}\quad\text{for all sufficiently large}\;L,
\end{IEEEeqnarray}
where \eqref{eq:dueto1} and \eqref{eq:dueto2} are due to $\frac{\gamma_Y}{\lambda_Y}=\frac{1-\rho_Y}{L\rho_Y}+O(\frac{1}{L^2})$.
Here, $\mathsf{D}_{\text{th},1}$ simplifies as follows:
\begin{IEEEeqnarray}{rCl}
\mathsf{D}_{\text{th},1}&=& \sigma_X^2-\sigma_X^4\gamma_Y^{-1}+O\left(\frac{1}{L}\right)\\
&=&d^{\infty}_{\min}+O\left(\frac{1}{L}\right).
\end{IEEEeqnarray}
So, for all $D\in (d_{\min}^{\infty},\sigma_X^2)$, the lower bound is given by $\underline{\mathcal{R}}^c(D)$ when $L$ is large enough. It just remains to approximate $\underline{\mathcal{R}}^c(D)$. Notice that  the second clause of \eqref{Rc-def} is active since
\begin{IEEEeqnarray}{rCl}
L\lambda_X^2(\lambda_Y-\gamma_Y)^{-1}+(L-1)\gamma_X^2(\gamma_X^{-1}-\gamma_Y^{-1})+\lambda_X &=& L(\sigma_X^2-\sigma_X^4\gamma_Y^{-1})+O(1)\\
&=& Ld_{\min}^{\infty}+O(1)<  LD.
\end{IEEEeqnarray}
The rate-distortion expression $\underline{\mathcal{R}}^c_2(D)$ can be  approximated as follows:
\begin{IEEEeqnarray}{rCl}
\underline{\mathcal{R}}^c_2(D)&=& 
\frac{L}{2}\log  \frac{(L-1)\gamma_X^2\gamma_Y^{-1}}{LD-\lambda_X-(L-1)(\gamma_X-\gamma_X^2\gamma_Y^{-1})}\\&=&\frac{L}{2}\log \frac{\sigma_X^4}{\gamma_YD-\sigma_X^2\gamma_Z}-\frac{1}{2}\frac{D-\sigma_X^2}{D-\sigma_X^2+\sigma_X^4\gamma_Y^{-1}}+O\left(\frac{1}{L}\right)\\
&=& \underline{\mathcal{R}}_2^{\infty}(D).
\end{IEEEeqnarray}
In summary, the analysis of this case yields~\eqref{rate-final-lower-4th-condition}. This concludes the proof.

\section{Conclusion}\label{sec:conclusion}
We have studied the problem of distributed compression of symmetrically correlated Gaussian sources. An explicit lower bound on the rate-distortion function is established and is shown to partially coincide with the Berger-Tung upper bound. The asymptotic expressions for the upper and lower bounds are derived in the large $L$ limit. It is of considerable theoretical interest to develop new bounding techniques to close the gap between the two bounds.



\appendices
\section{Sketch of Proof of Theorem~\ref{thm-ach}}\label{sketch-thm1-proof}

The proof is built upon the so-called Berger-Tung bound \cite[Thm 12.1]{ElGamal} as summarized in the following lemma.

\begin{lemma}\label{BT-lem} Let $\textbf{V}:=(V_1,\ldots,V_L)^T$ be an auxiliary random vector jointly distributed with $(\textbf{X},\textbf{Y},\textbf{Z})$ such that $(\textbf{X},\textbf{Z}, \{Y_{\ell'}\}_{\ell'\in\{1,\ldots,L\}\backslash \ell },\{V_{\ell'}\}_{\ell'\in\{1,\ldots,L\}\backslash \ell } )\to Y_{\ell}\to V_{\ell}$ form a Markov chain for $\ell=1,\ldots,L$. We have $R\geq \mathcal{R}(D)$ for any $(R,D)$ such that 
\begin{IEEEeqnarray}{rCl}
 R\geq  I(\textbf{Y};\textbf{V}),
\end{IEEEeqnarray}
and 
\begin{IEEEeqnarray}{rCl}
D\geq\frac{1}{L}\mathbb{E}[(\textbf{X}-\mathbb{E}[\textbf{X}|\textbf{V}])^T(\textbf{X}-\mathbb{E}[\textbf{X}|\textbf{V}])].
\end{IEEEeqnarray}
\end{lemma}
Let
$\textbf{Q} := (Q_1,\ldots,Q_L)^T$
be an $L$-dimensional zero-mean Gaussian random vector with covariance matrix
\begin{IEEEeqnarray}{rCl}
\Sigma_Q:= \text{diag}^{(L)}(\lambda_Q,\ldots,\lambda_Q),\label{Q-sig}
\end{IEEEeqnarray}
where $\lambda_Q>0$. We assume $\textbf{Q}$ is independent of $(\textbf{X},\textbf{Y},\textbf{Z})$.
Define the following auxiliary random variables:
\begin{IEEEeqnarray}{rCl}
V_{\ell} := X_{\ell} + Q_{\ell},\qquad \ell\in\{1,\ldots,L\}. \label{V-rv}
\end{IEEEeqnarray}
Note that the resulting $\textbf{V}$ satisfies the Markov chain constraints specified in Lemma~\ref{BT-lem}. One can readily complete the proof by verifying 
\begin{IEEEeqnarray}{rCl}
I(\textbf{Y};\textbf{V})= \frac{1}{2}\log \left(1+\frac{\lambda_Y}{\lambda_Q}\right)+\frac{L-1}{2}\log \left(1+\frac{\gamma_Y}{\lambda_Q}\right),\label{verify:ach-rate}
\end{IEEEeqnarray}
and
\begin{IEEEeqnarray}{rCl}
\mathbb{E}[(\textbf{X}-\mathbb{E}[\textbf{X}|\textbf{V}])^T(\textbf{X}-\mathbb{E}[\textbf{X}|\textbf{V}])]=\lambda_X\left(1-\frac{\lambda_X}{\lambda_Y+\lambda_Q}\right)+(L-1)\gamma_X\left(1-\frac{\gamma_X}{\gamma_Y+\lambda_Q}\right).\label{verify:distortion-ach}
\end{IEEEeqnarray}



\section{Sketch of Proof of Theorem~\ref{thm-conv}}\label{sketch-thm2-proof}

Let 
\begin{IEEEeqnarray}{rCl}(Y_1,\ldots,Y_{L})^T:=(U_1,\ldots,U_{L})^T+(W_1,\ldots,W_{L})^T,
\end{IEEEeqnarray}
where $(U_1,\ldots,U_{L})^T$ and $(W_1,\ldots,W_{L})^T$ are two mutually independent $L$-dimensional zero-mean Gaussian vectors with covariance matrices $\Sigma_U\succ 0$ and 
\begin{IEEEeqnarray}{rCl}
\Lambda_W:=\text{diag}^{(L)}(\lambda_W,\ldots,\lambda_W)\succ 0.\label{lw-def}
\end{IEEEeqnarray}
Then, two auxiliary random processes $\{(U_{1,i},\ldots,U_{L,i})^T\}_{i=1}^{\infty}$ and $\{(W_{1,i},\ldots,$ $W_{L,i})^T\}_{i=1}^{\infty}$ are constructed in an i.i.d. manner. 

According to Definition~\ref{model-def}, for any  $R\geq \mathcal{R}(D)$ and $\epsilon>0$, there exist encoding and decoding functions such that 
\begin{IEEEeqnarray}{rCl}
\frac{1}{n}\sum_{\ell=1}^L \log |\mathcal{M}_{\ell}|\leq R+\epsilon,
\end{IEEEeqnarray}
and 
\begin{IEEEeqnarray}{rCl}
\frac{1}{nL}\sum_{\ell=1}^L\sum_{i=1}^n \mathbb{E}[(X_{\ell,i}-\hat{X}_{\ell,i})^2]\leq D+\epsilon.\label{dist-cons2}
\end{IEEEeqnarray}
The proof is divided to several steps as follows.\\
\textbf{Simplifying the Rate Constraint}:
Lower bounding $\frac{1}{n}\sum_{\ell=1}^L \log|\mathcal{M}_{\ell}|$ by the standard information-theoretic arguments as in \cite[pp. 2349]{Jun} yields 
\begin{IEEEeqnarray}{rCl}
\frac{1}{2}\log \frac{\det(\Sigma_U)\det(\Lambda_W)}{\det(\Delta_{U|M})\det(\Delta_{Y|U,M})}\leq R+\epsilon,\label{rate-cons2}
\end{IEEEeqnarray}
where 
\begin{IEEEeqnarray}{rCl}
\Delta_{U|M}&:=& \frac{1}{n}\sum_{i=1}^n \mathbb{E}[(U_{j,i}-\hat{U}_{j,i})_{j\in \{1,\ldots,L\}}^T(U_{j,i}-\hat{U}_{j,i})_{j\in \{1,\ldots,L\}}],\label{DeltaUM-def}\\[-1ex]
\Delta_{Y|U,M}&:=& \frac{1}{n}\sum_{i=1}^n \mathbb{E}[(Y_{j,i}-\doublehat{Y}_{j,i})_{j\in \{1,\ldots,L\}}^T(Y_{j,i}-\doublehat{Y}_{j,i})_{j\in \{1,\ldots,L\}}],\label{DeltaYUM-def}
\end{IEEEeqnarray}
with
\begin{IEEEeqnarray}{rCl}
\hat{U}_{j,i}&:=&\mathbb{E}[U_{j,i}|(M_{\ell})_{\ell\in \{1,\ldots,L\}}],\\
\doublehat{Y}_{j,i}&:=& \mathbb{E}[Y_{j,i}|(U_{\ell}^n)_{\ell\in \{1,\ldots,L\}},(M_{\ell})_{\ell\in \{1,\ldots,L\}}].
\end{IEEEeqnarray}
We also define 
\begin{IEEEeqnarray}{rCl}
\delta_j := \sum_{i=1}^n\mathbb{E}[(Y_{j,i}-\bar{Y}_{j,i})^2],\qquad j\in \{1,\ldots,L\},
\end{IEEEeqnarray}
where 
\begin{IEEEeqnarray}{rCl}
\bar{Y}_{j,i}:=\mathbb{E}[Y_{j,i}|U_{j}^n,M_j].
\end{IEEEeqnarray}
It is clear that
\begin{IEEEeqnarray}{rCl}
\delta_j>0,\qquad j\in [1,L].\label{deltaj}
\end{IEEEeqnarray}
Furthermore, since $Y_{j}^n=U_j^n+W_j^n$, $j\in [1,L]$, and $(U_1^n,\ldots,U_L^n)$ and $(W_1^n,\ldots,W_L^n)$ are mutually independent, we have
\begin{IEEEeqnarray}{rCl}
\Delta_{Y|U,M}=\text{diag}^{(L)}(\delta_1,\ldots,\delta_L).\label{DeltaYUMdiag}
\end{IEEEeqnarray}
\textbf{Simplifying the Distortion Constraint:} We define
\begin{IEEEeqnarray}{rCl}
\Delta_{Y|M}&:=& \frac{1}{n}\sum_{i=1}^n \mathbb{E}[(Y_{j,i}-\hat{Y}_{j,i})_{j\in \{1,\ldots,
L\}}^T(Y_{j,i}-\hat{Y}_{j,i})_{j\in \{1,\ldots,L\}}],
\end{IEEEeqnarray}
where 
\begin{IEEEeqnarray}{rCl}
\hat{Y}_{j,i}&:=& \mathbb{E}[Y_{j,i}|(M_{\ell})_{\ell\in [1,L]}].\label{Yhat-def}
\end{IEEEeqnarray}
Clearly,
\begin{IEEEeqnarray}{rCl}
0\prec \Delta_{Y|M} \preceq \Sigma_Y.\label{cons5}
\end{IEEEeqnarray}
With some matrix calculations as in \cite[Appendix B]{Jun}, one can show that
\begin{IEEEeqnarray}{rCl}
\Delta_{U|M} &=& \Sigma_U \Sigma_Y^{-1}\Delta_{Y|M}\Sigma_Y^{-1}\Sigma_U+\Sigma_U-\Sigma_U \Sigma_Y^{-1}\Sigma_U,\label{DeltaUM}\\
\Delta_{Y|U,M} &\preceq& (\Delta_{Y|M}^{-1}+\Lambda_W^{-1}-\Sigma_Y^{-1})^{-1}.\label{dist-cons4}
\end{IEEEeqnarray}
Similar to $\Delta_{U|M}$ as in \eqref{DeltaUM}, one can show that
\begin{IEEEeqnarray}{rCl}
\frac{1}{n}\sum_{i=1}^n\sum_{j=1}^L \mathbb{E}[(X_{j,i}-\hat{X}_{j,i})^2] &= & \text{tr} (\Sigma_X \Sigma_Y^{-1}\Delta_{Y|M}\Sigma_Y^{-1}\Sigma_X+\Sigma_X-\Sigma_X \Sigma_Y^{-1}\Sigma_X).\label{simplified-dist}
\end{IEEEeqnarray}
Combining~\eqref{simplified-dist} and~\eqref{dist-cons2}, we get
\begin{IEEEeqnarray}{rCl}
\text{tr} (\Sigma_X\Sigma_Y^{-1}\Delta_{Y|M}\Sigma_Y^{-1}\Sigma_X+\Sigma_X-\Sigma_X \Sigma_Y^{-1}\Sigma_X)\leq L(D+\epsilon).\label{dist-cons3}
\end{IEEEeqnarray}
\textbf{Formulating the Optimization Problem:}
Considering~\eqref{rate-cons2}, \eqref{deltaj}, \eqref{DeltaYUMdiag}, \eqref{cons5}, \eqref{dist-cons4}, \eqref{dist-cons3}  and letting $\epsilon\to 0$, one can show using symmetrization and convexity arguments that there exist $\Delta$ with identical diagonal entries as well as identical off-diagonal entries and $\delta$ such that 
\begin{subequations}\label{opt-pr1}
\begin{IEEEeqnarray}{rCl}
&&\frac{1}{2}\log \frac{\det(\Sigma_U)}{\det(\Delta_{U|M})}+\frac{L}{2}\log \frac{\lambda_W}{\delta}\leq R,\label{obj100}\\
&&0\prec \Delta\preceq \Sigma_Y,\label{cons400}\\
&& 0<\delta ,\\
&&\text{diag}^{(L)}(\delta,\ldots,\delta)\preceq (\Delta^{-1}+\Lambda_W^{-1}-\Sigma_Y^{-1})^{-1},\\
&&\text{tr} (\Sigma_X \Sigma_Y^{-1}\Delta\Sigma_Y^{-1}\Sigma_X+\Sigma_X-\Sigma_X \Sigma_Y^{-1}\Sigma_X)\leq L D,\\
&& \Delta_{U|M} = \Sigma_U \Sigma_Y^{-1}\Delta\Sigma_Y^{-1}\Sigma_U+\Sigma_U-\Sigma_U \Sigma_Y^{-1}\Sigma_U.\label{cons401}
\end{IEEEeqnarray}
\end{subequations}
Using the eigenvalue decomposition, we have $\Delta = \Theta\;\text{diag}(\alpha,\beta,\ldots,\beta)\;\Theta^T$
for some positive $\alpha$ and $\beta$. So, inequality \eqref{obj100} can be equivalently written as
\begin{IEEEeqnarray}{rCl}
&&\frac{1}{2}\log \frac{\lambda_Y^2}{(\lambda_Y-\lambda_W)\alpha+\lambda_Y\lambda_W}+\frac{L-1}{2}\log \frac{\gamma_Y^2}{(\gamma_Y-\lambda_W)\beta+\gamma_Y\lambda_W}+\frac{L}{2}\log \frac{\lambda_W}{\delta}\leq R,\label{eq:lefthand}
\end{IEEEeqnarray}
and \eqref{cons400}--\eqref{cons401} reduce to the constraints \eqref{opt-consa}-\eqref{opt-consz}. Thus, minimizing the left-hand side of \eqref{eq:lefthand} over $(\alpha,\beta,\delta)$ subject to the constraints \eqref{opt-consa}-\eqref{opt-consz}  and
sending $\lambda_W$ to $\min(\lambda_Y,\gamma_Y)$ yields the desired lower bound.\vspace{-0.5cm}

\vspace{-0.5cm}

\bibliographystyle{IEEEtran}
\bibliography{references}

\end{document}